\pdfoutput=1
\documentclass[sigconf, authorversion]{acmart}
\usepackage{booktabs} 
\usepackage{subcaption}
\usepackage{algpseudocode}
\usepackage{algorithm}
\usepackage{color}
\usepackage{ wasysym }
\usepackage{graphicx}
\usepackage{listings}

\lstdefinelanguage{customC}{
  belowcaptionskip=1\baselineskip,
  breaklines=true,
  xleftmargin=\parindent,
  language=C,
  showstringspaces=false,
  basicstyle=\scriptsize\ttfamily,
  keywordstyle=\bfseries\color{green!40!black},
  commentstyle=\itshape\color{purple!40!black},
  identifierstyle=\color{blue},
  stringstyle=\color{orange},
  numbers=left,
  numbersep=5pt,
  numberstyle=\tiny\color{gray},
  escapechar=@
}

\definecolor{ibcolor}{rgb}{0.4,0.6,0.2}
\definecolor{clcolour}{rgb}{0.5,0.7,0.9}

\newcommand\lfb[0]{\textsc{FairFuzz}}

\newcommand\djpeg[0]{\texttt{djpeg}}
\newcommand\readpng[0]{\texttt{readpng}}
\newcommand\mutool[0]{\texttt{mutool}}
\newcommand\nm[0]{\texttt{nm}}
\newcommand\readelf[0]{\texttt{readelf}}
\newcommand\objdump[0]{\texttt{objdump}}
\newcommand\tcpdump[0]{\texttt{tcpdump}}
\newcommand\cxxfilt[0]{\texttt{c++filt}}
\newcommand\xmllint[0]{\texttt{xmllint}}

\newcommand\raritycutoff[0]{\textit{rarity\_cutoff}}
\newcommand\toolname[0]{\lfb{}}

\setcopyright{rightsretained}

\begin{document}
\sloppy
\title{\lfb{}: Targeting Rare Branches to Rapidly Increase Greybox Fuzz Testing Coverage}

\author{Caroline Lemieux, Koushik Sen}
\affiliation{%
  \institution{UC Berkeley}
}
\email{{clemieux,ksen}@cs.berkeley.edu}


\begin{abstract}
 In recent years, fuzz testing has proven itself to be one of the
 most effective techniques for finding correctness bugs and security   
 vulnerabilities in practice.  One particular fuzz testing tool, American Fuzzy
 Lop  or AFL, has become popular thanks to its ease-of-use
 and bug-finding power.  However, AFL remains limited in the depth of
 program coverage it achieves, in particular because it does not
 consider which parts of program inputs should not be mutated 
 in order to maintain deep program coverage.  We propose an    
  approach,  \lfb{}, that helps alleviate this limitation in
  two key steps. First, \lfb{} automatically prioritizes inputs
  exercising rare parts of the program under test. Second, it
  automatically adjusts the mutation of inputs so that the mutated
  inputs are more likely to exercise these same rare parts
  of the program.  We conduct evaluation on real-world programs against
  state-of-the-art versions of AFL, thoroughly repeating experiments to 
  get good measures of variability. We find that on certain benchmarks \lfb{} shows significant coverage increases after 24
  hours compared to state-of-the-art versions of AFL, while on others it achieves high program coverage at a significantly
  faster rate. 

\end{abstract}

\maketitle


\section{Introduction}

Fuzz testing has emerged as one of the most effective testing
techniques for finding correctness bugs and security vulnerabilities
in real-world software systems. It has been used successfully by major
software companies such as Microsoft~\cite{Godefroid08} and
Google~\cite{Evans11, Arya12, Moroz16} for security testing and
quality assurance. The term \emph{fuzz testing} is generally used to
designate techniques which test programs by generating random input data and
executing the program under such inputs. The goal of fuzz testing is
to exercise as many program paths as possible with the hope of
catching bugs that surface as assertion violations or program
crashes. While many individual test inputs generated by fuzz testing
may be garbage, due to its low computational overhead, fuzz testing
can generate test inputs much faster than more sophisticated methods
such as dynamic symbolic
execution~\cite{Godefroid05,Sen05,Cadar08,Sen06}. In practice, this
trade-off has paid off, and fuzz testers have found numerous
correctness bugs and security vulnerabilities in widely used
software~\cite{Stephens16, Bohme16,AFL,Hocevar07,Holler12,
  Householder12, Pacheco07, Yang11,
  Fraser11}. 

The success of one particular fuzz tester, American Fuzzy Lop (or
simply AFL)~\cite{AFL}, has gained attention both in practice and in
the research community~\cite{Stephens16, Li17, Bohme16, Bohme17}. This
fuzz tester alone has found vulnerabilities in a broad array of
programs, including Web browsers (e.g. Firefox, Internet Explorer),
network tools (e.g., tcpdump, wireshark), image processors (e.g.,
ImageMagick, libtiff), various system libraries (e.g., OpenSSH, PCRE),
C compilers (e.g., GCC, LLVM), and interpreters (for Perl, PHP,
JavaScript). A large part of its popularity can be attributed to its easy
setup. To use it, one needs only to compile it from source, compile
the program of interest with AFL's instrumented version of
\texttt{gcc} or \texttt{clang}, collect a sample input, and run
AFL.

Matching the simplicity of its setup is the simplicity of its
process. At its base, all AFL does is mutate the user-provided inputs
with byte-level operations. It runs the program under test on the
mutated inputs and collects some low-overhead program coverage
information from these runs. Then it saves the \emph{interesting} mutated inputs---the ones that
discover new program coverage. Finally, it continually repeats the process, 
starting with these interesting mutated inputs instead of the
user-provided ones.

While playing with AFL and its extensions, we observed---as others have in the past~\cite{Stephens16,Bohme16,Li17,laf-intel} --- one
 important limitation of AFL: AFL often fails to explore programs
very deeply, i.e. into sections guarded by specific constraints on the input. This is partly because AFL does not take into
consideration which parts of the input may be necessary for deeper program
exploration. For example, consider a program processing an image
format. This format may have a particular header indicating its validity,
and perhaps some special byte sequences which trigger different parts
of the image processor.  In order to cover deeper program functionality, 
an input must have this specific header and some of these
special byte sequences.  But even if AFL finds inputs with this
header or these sequences, it has no knowledge that the parts of
the inputs containing them are important for deep exploration of the
program.  Therefore, it is just as likely to mutate the header and special sequences as it is to mutate the data portion of the image. A bug which
only manifests when the header and sequences are in place may take a
much longer time to be generated than it would if these important
parts of the input were kept fixed.

We propose an approach, called \lfb{}, that tries to alleviate this
limitation.  \lfb{} functions in two main steps. First, it
automatically prioritizes inputs exercising rare parts of the program
under test. Second, it automatically adjusts byte-level mutation
techniques to increase the probability that mutated inputs will
exercise these same rare parts of the program while still exploring
different paths through the program. While we propose this mutation
modification strategy with the target of exercising rare parts of the
program, it could be adapted to target other interesting parts of the 
program, e.g., a recently patched function.

We created an open-source implementation of \lfb{} on top of AFL. We evaluated
\lfb{} against three popular versions of AFL: AFL,
FidgetyAFL~\cite{FidgetyAFL}) (a modification of AFL to better match
the behavior of AFLFast~\cite{Bohme16}), and
AFLFast.new~\cite{AFLFast.new} (an enhancement of AFLFast to match the behavior of FidgetyAFL).

We evaluated our approach on four general fronts. First, whether it
would result in faster program coverage. Relatedly, whether it would
result in more extensive coverage after a standard time budget (24
hours). Third, whether the modification of AFL's mutation
technique results in a higher proportion of inputs exercising the
targeted parts of the program. Finally, whether
it would enhance AFL's crash-finding ability. We conduct 
our evaluation on nine real-world benchmarks, including those 
used to evaluate AFLFast. We repeat our experiments and provide
measures of variability when comparing techniques. We find that on certain benchmarks \toolname{} leads to
significant coverage increases after 24 hours compared to AFL,
FidgetyAFL, and AFLFast.new, and that on other benchmarks \lfb{} achieves wide
program coverage at a significantly faster rate. We also observed that \lfb{} tends to perform better on programs with many nested conditionals. 

In summary, we make the following contributions:
\begin{itemize}
\item We propose an approach to increase the coverage obtained by
  greybox fuzzers by targeting rare branches, including a more general
  method to target mutations to a program characteristic of interest.
\item We develop an open-source\footnote{\url{https://github.com/carolemieux/afl-rb}} implementation of this approach built on top of
  AFL, named \textit{\lfb{}}.
\item We perform evaluation of \lfb{} against different
  state-of-the-art versions of AFL on real-world benchmarks with
  numerical information presented with a \emph{measure of variability}
  (i.e. confidence intervals).
\item When comparing state-of-the-art versions of AFL, we
  repeat our experiments 20 times to obtain statistically significant
  results and to measure variability in the outcomes.  We show that
  repetition and variability measures are necessary to draw correct
  conclusions while comparing AFL-based techniques due to the highly
  non-deterministic nature of AFL.  To our knowledge, we are the first
  work on AFL-based techniques to report a measure of variability in
  the coverage the techniques achieve through time.
\end{itemize}

We detail the general method and implementation on top of AFL in
Section~\ref{sec:method} and the performance results in
Section~\ref{sec:eval}.  We will begin in Section~\ref{sec:overview}
with a more detailed overview of AFL, its current limitations, and how
to overcome these with our method.


\section{Overview}

\label{sec:overview}

Our proposed technique, \lfb{}, is built on American Fuzzy Lop
(AFL)~\cite{AFL}.  AFL is a popular greybox mutation-based fuzz
tester. 
\footnote{\emph{Greybox} fuzz testers~\cite{AFL, libFuzzer} are designated as such since, unlike
\emph{whitebox} fuzz testers~\cite{Godefroid05,Sen06,Cadar08,Godefroid08}, they do
not do any source code analysis, but, unlike pure \emph{blackbox} fuzz
testers~\cite{Hocevar07}, they use limited
feedback from the program under test to guide their fuzzing strategy.}
Next we give a brief description of AFL, illustrate one of its limitations,
and motivate the need for \lfb{}.


To fuzz test programs, AFL generates random inputs. However, instead
of generating these inputs blindly, from scratch, it selects a set
of previously generated inputs and mutates them to derive new random
inputs.  A key innovation behind AFL is its use of coverage
information collected during the execution of the program on its previously-generated inputs. Specifically, AFL uses this information to select inputs 
for mutation, selecting only those that have achieved new program coverage.  
In order to collect this coverage information efficiently,
AFL inserts instrumentation into the program under test. To track
coverage, it first associates each basic block with a random number
via instrumentation.  The random number is treated as the \emph{unique
  ID} of the basic block.  The basic block IDs are then used to
generate unique IDs for the transitions between pairs of basic
blocks. In particular, for a transition from basic block $A$ to $B$,
AFL uses the IDs of each basic block---$ID(A)$ and $ID(B)$,
respectively---to define the ID of the transition, $ID(A\to B)$, as
follows:
$$ ID(A\to B)=(ID(A) \gg 1) \oplus ID(B). $$
Right-shifting ($\gg$) the basic block ID of the transition start
block ($A$) ensures that the transition from $A$ to $B$ has a
different ID from the transition from $B$ to $A$. We associate the
notion of basic block transition with that of a \emph{branch} in the
program's control flow graph, and throughout the paper we will use the
term branch to refer to this AFL-defined basic block transition unless
stated otherwise.  Note that since random numbers are used as unique
IDs for basic blocks, there is a small but non-zero probability
of having the same ID for two different branches. However, AFL's
creator argues~\cite{Zalewski17} that for many programs the actual
branch ID collision rate is small.

The coverage of the program under test on a given input is collected
as a set of pairs of the form (\textit{branch ID}, \textit{hit
  count}).  If a (\textit{branch ID}, \textit{hit count}) pair is
present in the coverage set, it denotes that during the execution of the
program on the input, the branch with ID \textit{branch ID} was exercised
\textit{hit count} number of times.  The hit count is bucketized to
small powers of two. 
AFL refers this set of pairs as the \emph{path} of an input. AFL
says that an input achieves \emph{new coverage} if it hits a new
branch (one not exercised in any previous execution) or achieves
a new hit count for an already-exercised branch---i.e. if it discovers
a new (\textit{branch ID}, \textit{hit count}) pair. 

\begin{algorithm}[t]
\caption{AFL Fuzzing}\label{alg:afl-outline}
\begin{algorithmic}[1]
\Procedure{FuzzTest}{$Prog,Seeds$}
\State $Queue \gets Seeds$ \label{line:afl-outline-init-queue}
\While{true} \Comment{begin a queue cycle} \label{line:afl-outline-begin-cycle}
\For{$seed$ in $Queue$} \label{line:afl-outline-traverse-queue}
\If{$\neg$\Call{isWorthFuzzing}{$seed$}}  \label{line:afl-outline-worth-fuzzing}
\State continue
\EndIf
\For{$i$ in $0$ to \Call{length}{$seed$}}  \label{line:afl-outline-det-begin}
\State $newinput \gets $\Call{mutateDeterministic}{$seed$, i}  \label{line:afl-outline-mutate-det}
\State \Call{runAndMaybeSave}{$Prog,newinput, Queue$}
\EndFor \label{line:afl-outline-det-end}
\State $score \gets $\Call{performanceScore}{$Prog,seed$} \label{line:afl-outline-havoc-begin}
\For{$i$ in $0$ to $score$} 
\State $newinput \gets $\Call{mutateHavoc}{$seed$}  \label{line:afl-outline-mutate-havoc}
\State \Call{runAndMaybeSave}{$Prog,newinput, Queue$}
\EndFor \label{line:afl-outline-havoc-end}
\EndFor
\EndWhile
\EndProcedure
\Procedure{runAndMaybeSave}{$Prog, input, Queue$} 
\State $runResults \gets$ \Call{run}{$Prog,input$} \label{line:afl-outline-run-program}
\If {\Call{newCoverage}{$runResults$}}
\State \Call{addToQueue}{$input, Queue$} \label{line:afl-outline-save-if-interesting}
\EndIf
\EndProcedure
\end{algorithmic}
\end{algorithm}

The overall AFL fuzzing algorithm is given in Algorithm~\ref{alg:afl-outline}. The fuzzing routine takes as input a
program and a set of user-provided \emph{seed inputs}. The seed inputs
are used to initialized a \emph{queue} (Line~\ref{line:afl-outline-init-queue}) of
inputs.  The queue contains the inputs which AFL will mutate in order to generate new inputs.  AFL goes through this queue
(Line~\ref{line:afl-outline-traverse-queue}), selects an input to
mutate (Line~\ref{line:afl-outline-worth-fuzzing}), mutates the input
(Lines~\ref{line:afl-outline-mutate-det},~\ref{line:afl-outline-mutate-havoc}),
runs the program on and, simultaneously, collects the coverage
information for the mutated inputs (Line~\ref{line:afl-outline-run-program}),
and finally adds these mutated inputs to the queue if they achieve new
coverage (Line~\ref{line:afl-outline-save-if-interesting}). An entire
pass through the queue is called a \emph{cycle}; cycles are repeated
(Line~\ref{line:afl-outline-begin-cycle}) until the fuzz testing
procedure is stopped by the user.

AFL's mutation strategies assume the input to the program under test is a
sequence of bytes, and can be treated as such during mutation. 
AFL mutates inputs under two main stages: the
\emph{deterministic} (Algorithm~\ref{alg:afl-outline},
Lines~\ref{line:afl-outline-det-begin}-\ref{line:afl-outline-det-end})
stages and the \emph{havoc}
(Lines~\ref{line:afl-outline-havoc-begin}-\ref{line:afl-outline-havoc-end})
stage. 
All the deterministic mutation stages operate by traversing the input 
under mutation and applying a mutation at sequential positions (bits and bytes)
in this input. These mutations includes bit flipping, byte flipping,
arithmetic increment and decrement of byte values, replacing of bytes
with ``interesting'' values, etc. The number of mutated inputs
produced in each of these stages is governed by the length of the 
input being mutated.  On the other hand, the havoc stage works by 
applying a sequence of random mutations (e.g. setting random bytes 
to random values, deleting or cloning subsequences of the input) to
the input being mutated to produce a new input. 
The number of total havoc-mutated inputs to be produced is determined
by a performance score, \textit{score}
(Line~\ref{line:afl-outline-havoc-begin}). 

AFL's mutation strategies pay little or no attention to the contents
of an input.  Therefore, they can easily destroy parts of an existing
input that are necessary to explore deeper parts of a program.  To see
how this could make it difficult for AFL to explore deeper parts
of the program under test, consider the code in
Figure~\ref{fig:xmllint-ex}. This fragment, simplified from a portion
of the \texttt{libxml} file \texttt{parser.c}, shows the various
nested string comparisons that process XML default
declarations. Exploring the correct handling of, say, the requirement
that the \texttt{\#FIXED} keyword must be followed by a blank
character (Line 16) requires producing an input containing the string 
\texttt{<!ATTLIST}, followed by a correct attribute type string like
\texttt{ID} or \texttt{CDATA} (to pass Line 3), then finally the
string \texttt{\#FIXED}. If AFL manages to produces the input
\texttt{<!ATTLIST BD}, it will not prioritize mutation of the bytes
after \texttt{<!ATTLIST}, and is as likely to produce the mutants
\texttt{<!CATLIST BD}, \texttt{<!!ATTLIST BD}, \texttt{???!ATTLIST BD}
as it is to produce \texttt{<!ATTLIST ID}.\footnote{AFL does perform
  some more complex mutations only on positions which, when mutated,
  cause different program behavior. In this case, \texttt{<!CATLIST
    BD}, \texttt{<!!ATTLIST BD}, and \texttt{???!ATTLIST BD} are
  generated by mutations performed at locations whose modification
  results in different behavior.}  However, to explore the code in
Figure 1 more rapidly, once \texttt{<!ATTLIST BD} has been discovered,
mutations should keep the \texttt{<!ATTLIST } part of this input
constant.

\begin{figure}
\begin{lstlisting}[language=customC]
if (CMP9(ptr,'<','!','A','T','T','L','I','S','T')) {
   // if an attribute type is found, process default val 
    if (xmlParseAttributeType(ptr) > 0) { 
       if (CMP9(ptr,'#','R','E','Q','U','I','R','E','D')) {
          ptr += 9;
          default_decl = XML_ATTRIBUTE_REQUIRED;
       }
     if (CMP8(ptr,'#','I','M','P','L','I','E','D')) {
         ptr += 8;
         default_decl = XML_ATTRIBUTE_IMPLIED;
      }
      if (CMP6(ptr,'#','F','I','X','E','D')) {
        ptr += 6;
        default_decl = XML_ATTRIBUTE_FIXED;
        if (!IS_BLANK_CH(ptr)) {
           xmlFatalErrorMsg("Space required after '#FIXED'");
         }
      }
   }
}
\end{lstlisting}
\caption{Code fragment based off the \texttt{libxml} file \texttt{parser.c} showing many nested if statements that must be satisfied to explore erroneous behavior.}\label{fig:xmllint-ex}
\end{figure}


We propose a two-pronged approach that builds on AFL to do exactly
this. The first part of our approach is the identification of
statements like the if statements in Figure~\ref{fig:xmllint-ex} (and
Line 3 of Figure~\ref{fig:magic-number}). For this, we leverage the
fact that such statements are indeed hit by very few of AFL's generated
inputs (i.e. they are \emph{rare}), and can thus be easily identified
by keeping the track of the number of inputs which hit each
branch. 
Having identified these rare branches for targeted fuzzing, we modify
the input mutation strategy in order to keep the condition of the rare
branch satisfied.  Specifically, we use a deterministic mutation phase
to approximately determine the parts of the input that cannot be
mutated if we want the input to hit the rare branch.  The subsequent 
mutation stages are then not allowed to mutate these crucial parts
of the input.  As a result, we significantly increase the probability
of generating new inputs that hit the rare branch.  This opens up the possibility of better exploring the part
of the code that is guarded by the if condition.  We find this
approach leads to significant coverage increases after 24 hours
compared to stock AFL and other modified versions of AFL on some
benchmarks, and that on other benchmarks it achieves wide program
coverage at a significantly faster rate. We will present the details
of this approach in the next
section. 


We note that prior work on AFL has focused on a closely related, but
not identical, issue. This issue is usually illustrated with code
fragments like that in Figure~\ref{fig:magic-number}. In this
fragment, a program crash hides behind a branch statement that is only
satisfied by the presence of a certain \emph{magic number} in the
input: AFL's mutation-based strategy is highly unlikely to discover
the 8 consecutive magic bytes. Several techniques have been proposed
to allow AFL to better explore scenarios like
these. AFLFast~\cite{Bohme16} assumes these will present as rare paths
and applies more havoc mutations on seeds exercising rare paths to try
and hit more rare paths; Driller~\cite{Stephens16} uses symbolic
execution to produce these magic numbers when AFL gets stuck;
Steelix~\cite{Li17} adds a static analysis stage, extra
instrumentation, and mutations to AFL to exhaustively search the
bytes around a byte which is matched in a multi-byte comparison. The
issue of producing a single input with a magic number is fundamentally
orthogonal to the issue of making sure this magic number retains its
value. None of these techniques will be able to prevent this magic
number from being mutated during AFL's usual mutation stages after it
has been discovered, while our technique does. Additionally, since
Steelix does not instrument one byte comparison instructions, it would
not effectively explore the code in Figure~\ref{fig:xmllint-ex}, as
the \texttt{CMP}$n$ macros compile into single byte comparison
instructions. Since AFLFast's emphasis on rare paths is similar to our
emphasis on rare branches, we will evaluate our technique against
AFLFast. 


\begin{figure}
\begin{lstlisting}[language=customC]
int main( int argc, const char* argv[] ) {
// ...
   if (magic == 0xBAAAAAAD) {
     // crash!   
   }
}
\end{lstlisting}
\caption{Magic number guarding a crash.}\label{fig:magic-number}
\end{figure}


\section{\lfb{} Algorithm}
\label{sec:method}
%
In \lfb{}\footnote{\lfb{} is so called because it gives more priority to the rare
branches of a program, which are ``unfairly'' not prioritized by stock AFL.}, we modify the AFL algorithm in two key ways  to target 
exploration to rare branches. First, we modify the selection of inputs
to mutate from the queue (\textsc{isWorthFuzzing} in
Algorithm~\ref{alg:afl-outline},
Line~\ref{line:afl-outline-worth-fuzzing}) in order to select inputs
which hit rare branches. Second, we modify the mutations that are
performed on these inputs (\textsc{mutateDeterministic} on
Line~\ref{line:afl-outline-mutate-det} and \textsc{mutateHavoc} on
Line~\ref{line:afl-outline-mutate-havoc}) in order to increase the
probability that the mutated inputs will hit the rare branch in
question. We describe the modifications to input selection in
Section~\ref{sec:queue-modification} and the modifications to the
mutation strategies in Section~\ref{sec:mutation-modification}.


\subsection{Selecting Inputs to Mutate}
\label{sec:queue-modification}


Recall from Section~\ref{sec:overview} that conventional AFL operates
by repeatedly traversing a queue of inputs and mutating these inputs
to produce new inputs. Unlike a traditional queue data structure, 
inputs are never truly removed from the queue. Instead, the
method \textsc{isWorthFuzzing} selects certain inputs to mutate in
each cycle. The method is non-deterministic, prioritizing short, fast,
and recently discovered inputs, but sometimes selecting old
inputs for mutation. We replace \textsc{isWorthFuzzing} with the
function \textsc{hitsRareBranch}, which, as the name suggests, returns
true if the input hits a rare branch and false otherwise.

In order to do this, we must define what it means to be a rare
branch. A natural idea is to designate the $n$ branches hit by the
fewest inputs as rare, or the branches hit by less than $p$ percent of
inputs to be rare.  After some initial experiments with our benchmark
programs, we rejected these
methods as (a) they can fail to capture what it means to be rare
(e.g. if $n=5$ and the two rarest branches are hit by 20 and 15,000
inputs, both would be equally rare when this is clearly not the case),
and (b) these thresholds would likely
need to be modified for different
benchmarks.

Instead, we say a branch is \emph{rare} if it has been hit by a number
of inputs less than or equal to a dynamically chosen $\raritycutoff$.
Informally, $\raritycutoff$ is the smallest power of two which bounds
the number of inputs hitting the rarest branch. For example, if the
rarest branch has been hit by only 17 inputs, we would cut off rarity
at branches that have been hit by $\leq 2^5$ inputs. Formally, if
$B_\text{hit}$ is the set containing the number of inputs hitting each
branch,
$$ \raritycutoff = 2^i \text{ where } 2^{i-1} < min(B_\text{hit}) \leq 2^{i}.$$
\lfb{} keeps track of $B_\text{hit}$ by keeping a map of branch IDs to the
number of inputs which hit the corresponding branch so far (the
\emph{input count}). After running the program on an input, \lfb{}
increments the input count by one for each branch the input hits. \lfb{}
calculates \raritycutoff{} at every call of
\textsc{hitsRareBranch}. 


During fuzzing, \lfb{} goes over the queue and selects for mutation
inputs on which \textsc{hitsRareBranch} returns true.  Note that the
execution of a program on a selected input may hit multiple rare
branches.  In that case, \lfb{} picks the rarest branch among them as the
\emph{target branch} for the purpose of mutation, and subsequently tries
to make sure that mutations of the selected input hit this target
rare branch more frequently. Of course, if  the input hits only one rare branch,
this is the target branch.
Since at the beginning of fuzz testing no inputs have been produced,
and the definition of rare requires some number of inputs to have hit
a branch, we run a round of regular AFL mutation on the user-provided
input. All subsequent inputs are selected from the queue with
\textsc{hitsRareBranch}.

Finally, due to its strict boolean nature, \textsc{hitsRareBranch} can
get stuck
more easily than the probabilistic \textsc{isWorthFuzzing} (i.e. by
repeatedly pulling the same input from the
queue). Currently, we take action only in the extreme case, when none of the
mutated inputs produced from a seed hit the target
branch. If we witness this behavior, we put the rare branch on an
\emph{exclude list}. 
We ignore any branches in the exclude list when finding rare branches
and calculating the rare branch cutoff in the future.


\subsection{Mutating Inputs}
\label{sec:mutation-modification}

After selecting an input which hits a rare branch, we bias mutation of
this input to produce inputs which hit the target branch at a higher
frequency than AFL would. The biggest part of this is the calculation
of a \emph{branch mask}, which we use to influence all subsequent
mutations.

The branch mask designates at which positions in the input can
bytes be (1) overwritten (\emph{overwritable}), (2) deleted
(\emph{deletable}), or (3) inserted (\emph{insertable}) while the
resulting input still hits the target branch.  We compute this branch
mask in AFL's deterministic mutation stages. Part (1) of the branch
mask is computed in the byte flipping stage: if the input resulting
from a byte flip still hits the target branch, the position is
designated as overwritable. We add two new deterministic stages after
the byte flipping to compute parts (2) and (3) of the mask. First, we
traverse the input and delete each byte in sequence, marking the
position as deletable if its removal does not make the program miss
the target  branch. Similarly, we traverse the input and add a random byte
before each position (and at the end of the input), marking the
position as insertable if the resulting input still makes the program
hit the target  branch.  Note that this technique for
computing the branch mask is approximate---even if \lfb{} determines
that a position in the input is overwritable, a particular mutation of the byte
at that position might make the program miss the target branch, or vice-versa.
However, in our experiments we found that the use of branch mask
significantly increases the probability of hitting the target branch (see Section~\ref{sec:eval-targeting-branches}).

After its calculation, the branch mask is used to influence all subsequent mutations of the input. 
During the deterministic stages, a mutation is performed at a position only if the branch mask
indicates the target branch can still be hit by the resulting input. 
In the havoc stage, instead of performing  random mutations 
at randomly selected positions in the input, \lfb{} performs 
random mutations at positions randomly selected \emph{within the
  corresponding modifiable positions} of the branch
mask.  
Note that when, during havoc mutations, bytes are deleted from or added to the mutated
input, the branch mask created for the input at the beginning of the mutation 
process no longer maps to the input being mutated. As such,  \lfb{} modifies
the branch
mask 
in coordination with the mutated input. When a block of the input is
deleted,  \lfb{} deletes the corresponding block within the branch mask;
when a block is inserted, \lfb{} inserts a section into the branch mask
which designates the new area as completely modifiable. Of course, this
is an approximation, so we expect the proportion of inputs hitting
the target branch in the havoc stage to be smaller than the proportion of
inputs hitting the rare branch in deterministic fuzzing stages. We
examine this further in Section~\ref{sec:eval-targeting-branches}.



\subsection{Trimming Inputs for Target Branches}

AFL's efficiency depends on large part on its ability to quickly produce and modify
inputs~\cite{Zalewski17}.  Thus, it is important to make sure \lfb{}'s branch mask
computation is efficient. Since the runtime of the branch mask computation is linear 
in the length of the selected input, \lfb{} needs to keep the
length of the inputs in the queue short. AFL has two techniques
for keeping inputs short: (1) prioritizing short inputs when selecting
inputs for mutation and (2)  trimming  (i.e. performing an efficient approximation of
delta-debugging~\cite{Zeller02}) the input it selects for mutation
before mutating it. 
Trimming
attempts to minimize the input selected for mutation with the constraint that the minimized
input hits the same path (set of (\textit{branch ID}, \textit{hit
  count}) pairs) as the seed input.  However, this constraint
is not good enough for reducing the length of inputs significantly when
very long inputs are chosen---something \lfb{} may do since it selects
inputs only based on whether they hit a rare branch. We found that we 
can make inputs shorter in spite of this if we relax the trimming
constraint to require that the minimized input hits only the target branch
of the original input, instead of the same path as the original input.  
We refer to our technique with
this relaxed constraint as our technique ``with
trimming''.
\section{Implementation and Evaluation}
\label{sec:eval}

We have implemented our technique as \emph{an open-source extension} of AFL named \lfb{}. This implementation adds around 600 lines of C code to the file
containing AFL's core implementation, including some code used only
for experimentation, i.e. for the statistics in 
Section~\ref{sec:eval-targeting-branches}. 


\begin{figure*}[t]
\centering
\begin{subfigure}[t]{0.66\columnwidth}
\subcaption{tcpdump} \label{fig:coverage-tcpdump}
\includegraphics[width=\columnwidth]{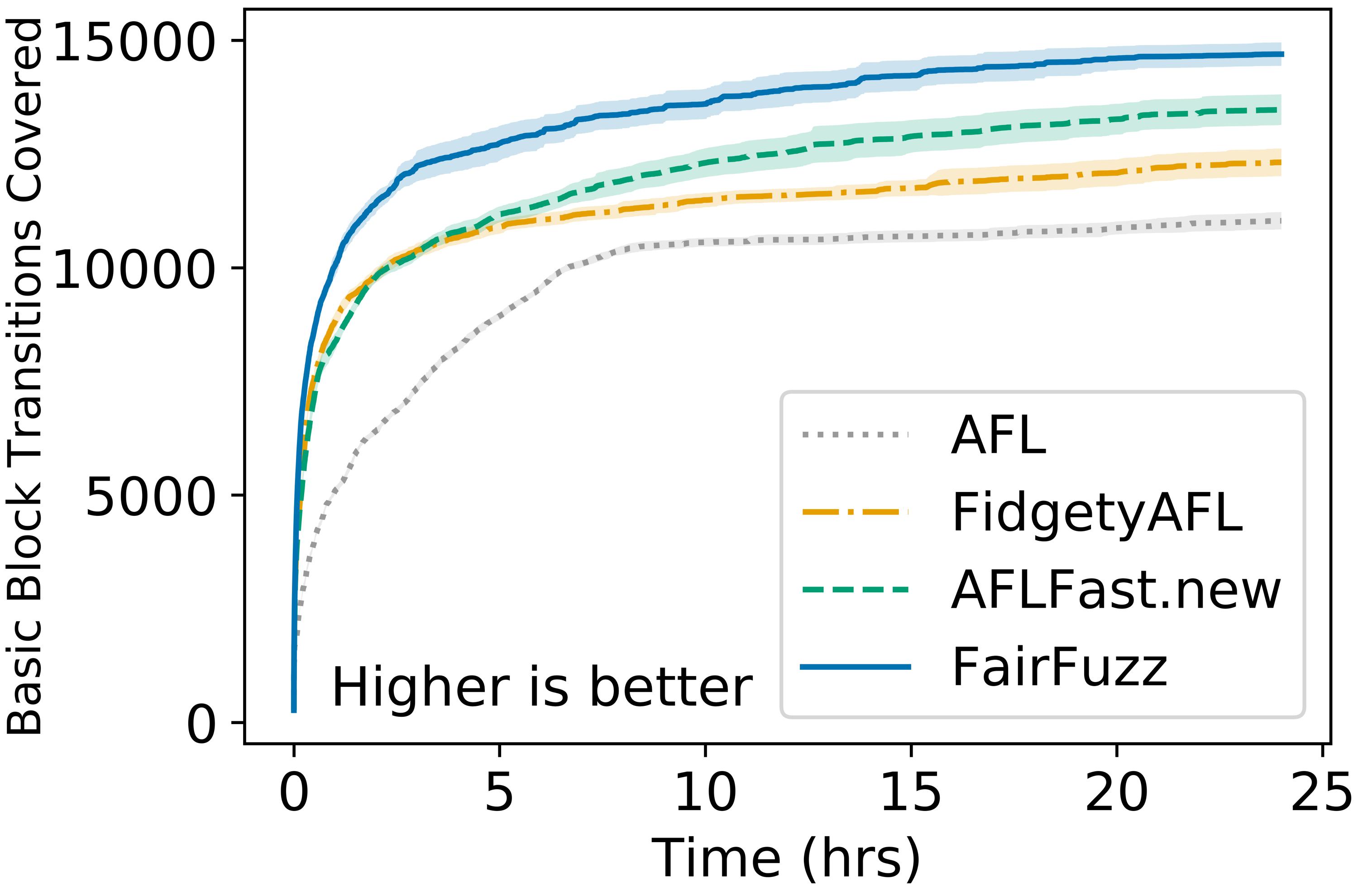}
\end{subfigure}
\begin{subfigure}[t]{0.66\columnwidth}
\subcaption{readelf}\label{fig:coverage-readelf}
\includegraphics[width=\columnwidth]{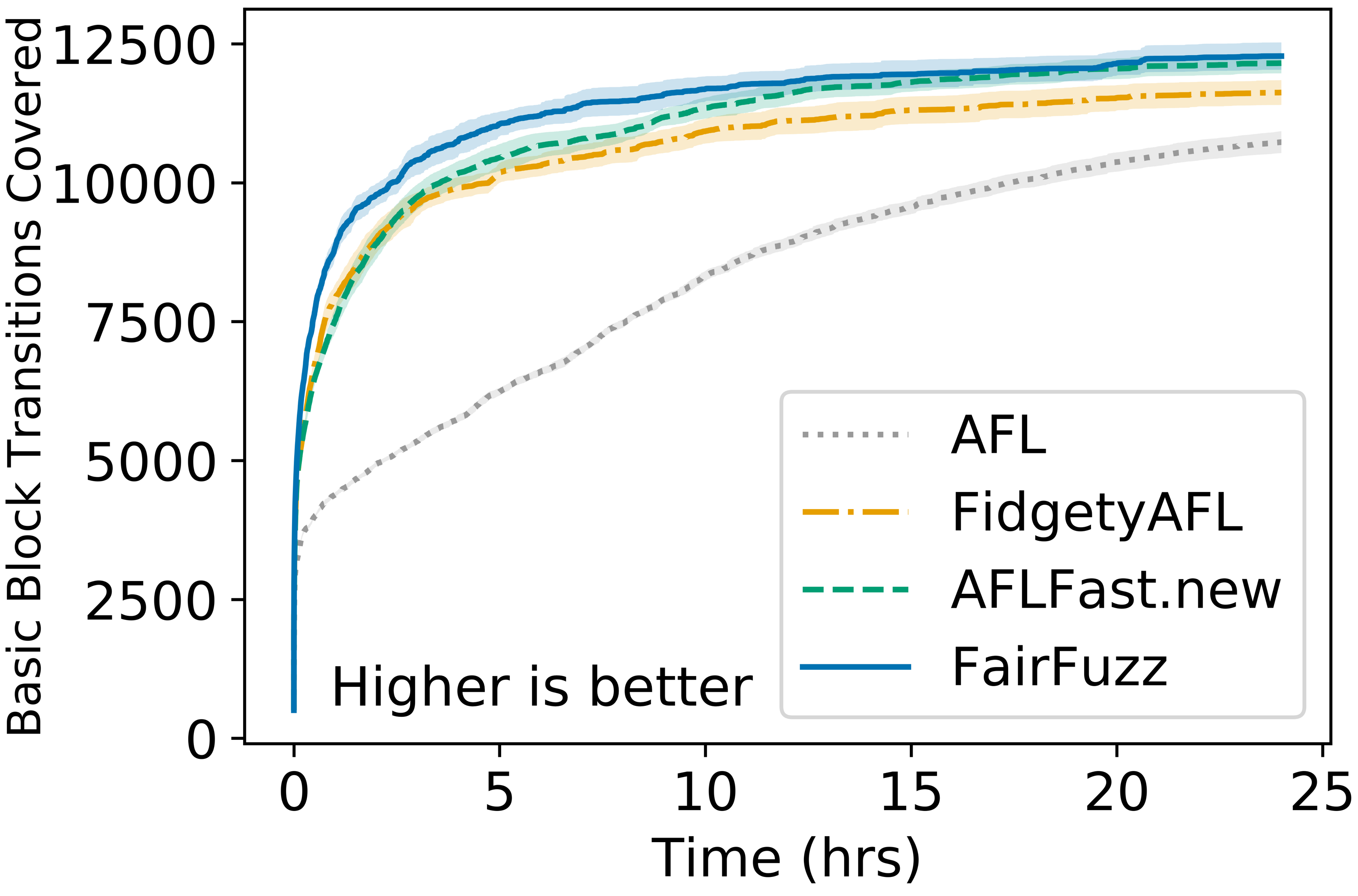}
\end{subfigure}
\begin{subfigure}[t]{0.66\columnwidth}
\subcaption{nm}\label{fig:coverage-nm}
\includegraphics[width=\columnwidth]{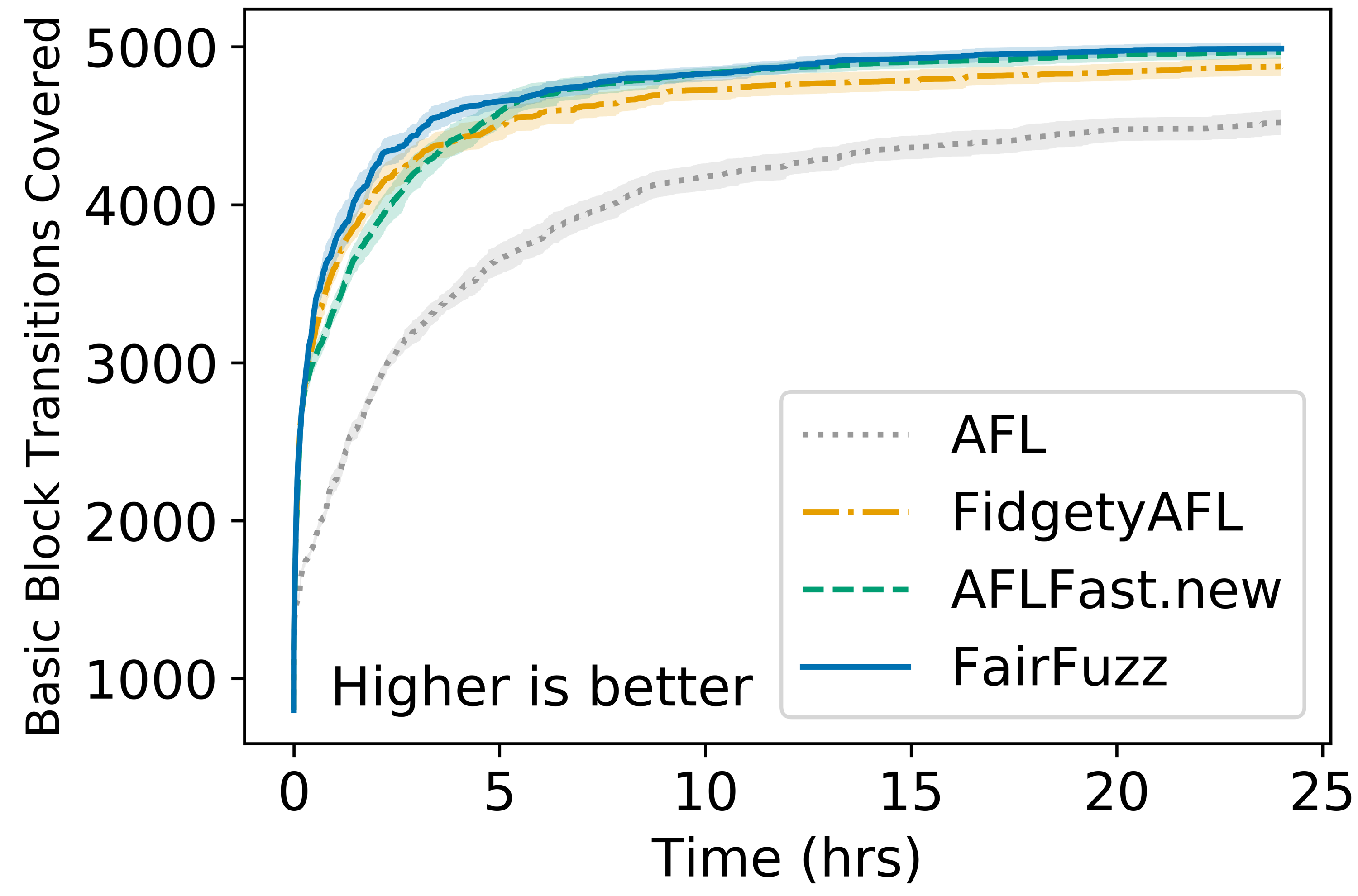}
\end{subfigure}
\begin{subfigure}[t]{0.66\columnwidth}
\subcaption{objdump}\label{fig:coverage-objdump}
\includegraphics[width=\columnwidth]{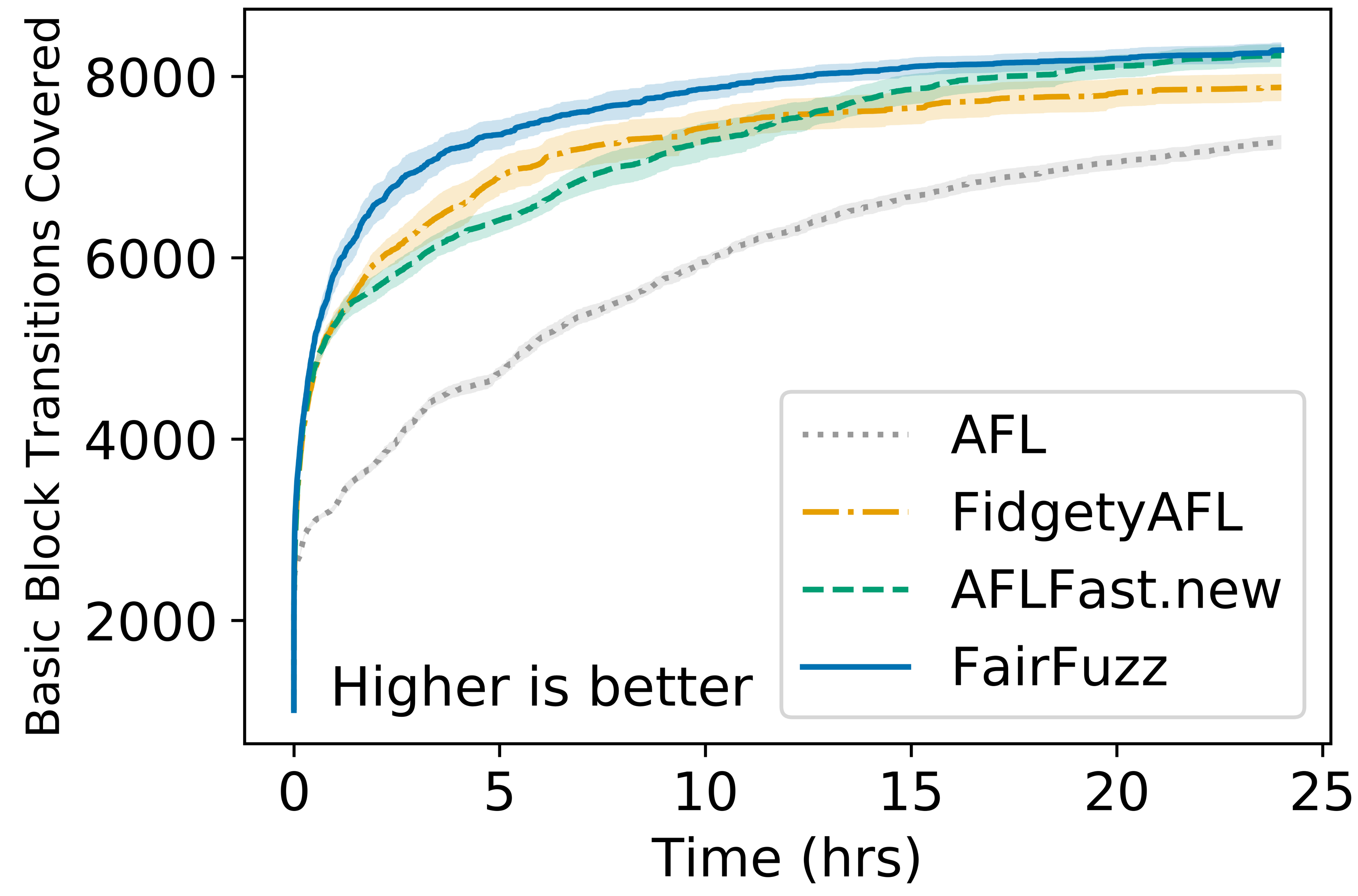}
\end{subfigure}
\begin{subfigure}[t]{0.66\columnwidth}
\subcaption{c++filt}\label{fig:coverage-cxxfilt}
\includegraphics[width=\columnwidth]{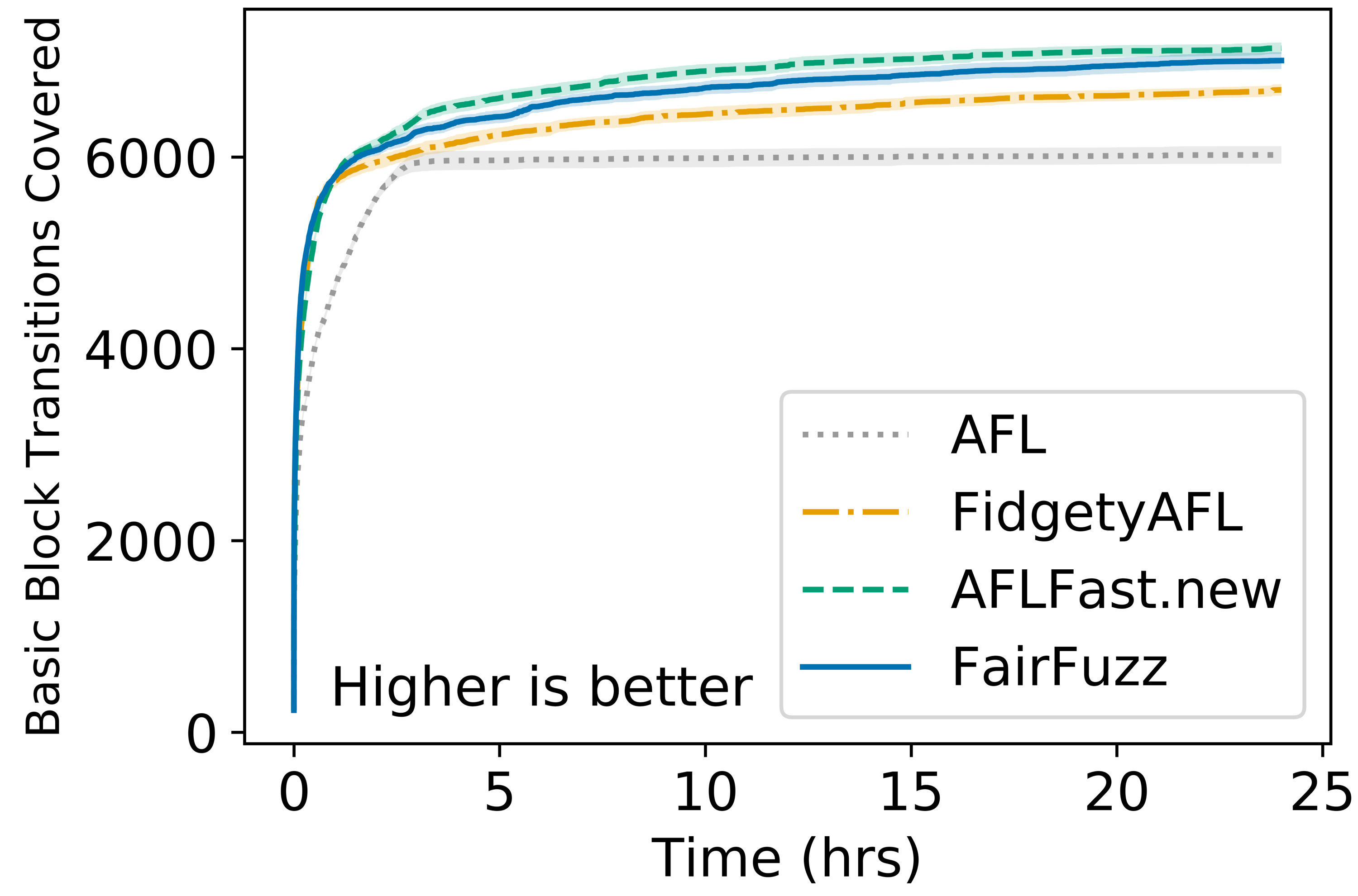}
\end{subfigure}
\begin{subfigure}[t]{0.66\columnwidth}
\subcaption{xmllint}\label{fig:coverage-xmllint}
\includegraphics[width=\columnwidth]{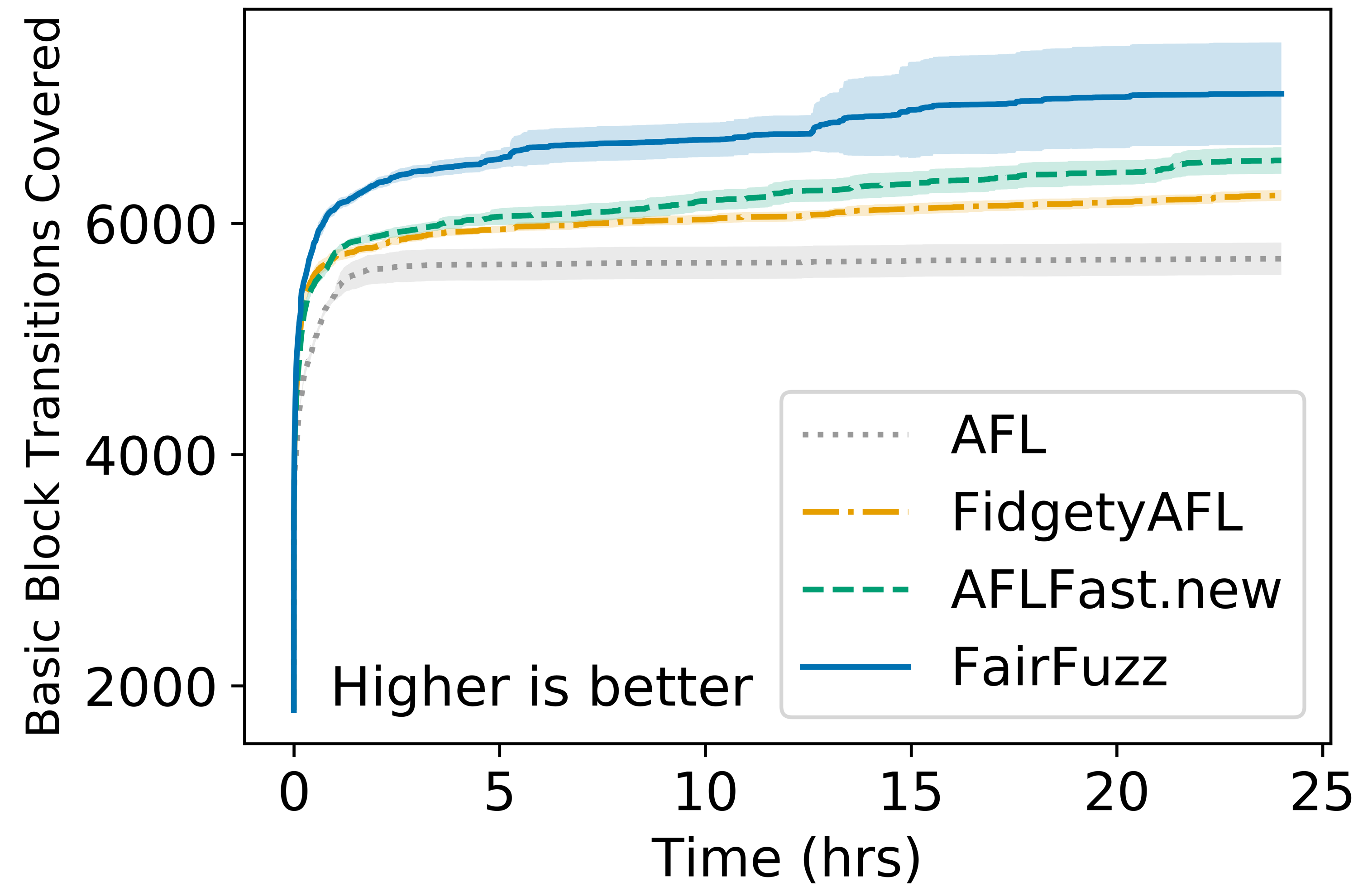}
\end{subfigure}
\begin{subfigure}[t]{0.66\columnwidth}
\subcaption{mutool draw}\label{fig:coverage-mutool}
\includegraphics[width=\columnwidth]{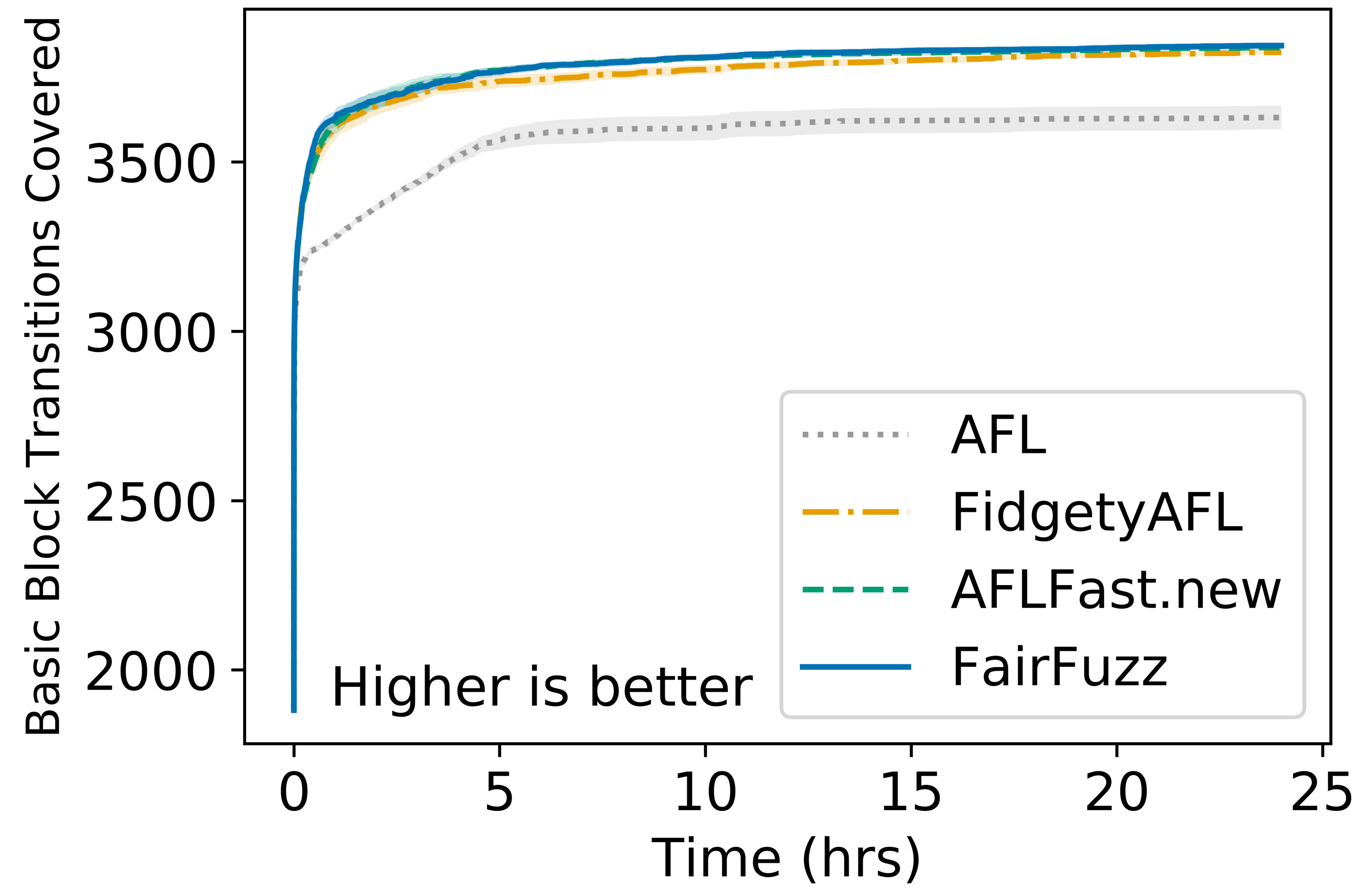}
\end{subfigure}
\begin{subfigure}[t]{0.66\columnwidth}
\subcaption{djpeg}\label{fig:coverage-djpeg}
\includegraphics[width=\columnwidth]{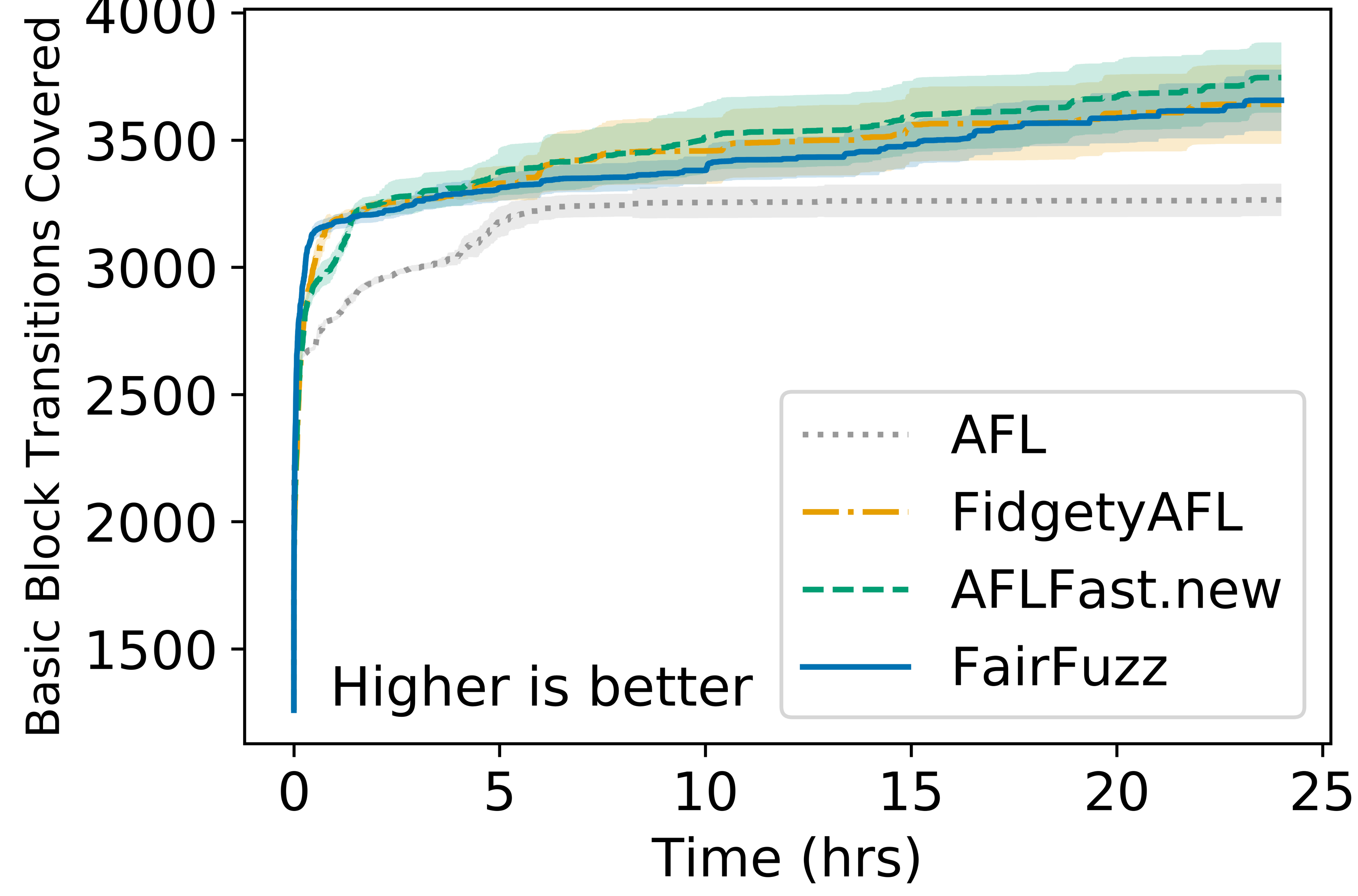}
\end{subfigure}
\begin{subfigure}[t]{0.66\columnwidth}
\subcaption{readpng}\label{fig:coverage-readpng}
\includegraphics[width=\columnwidth]{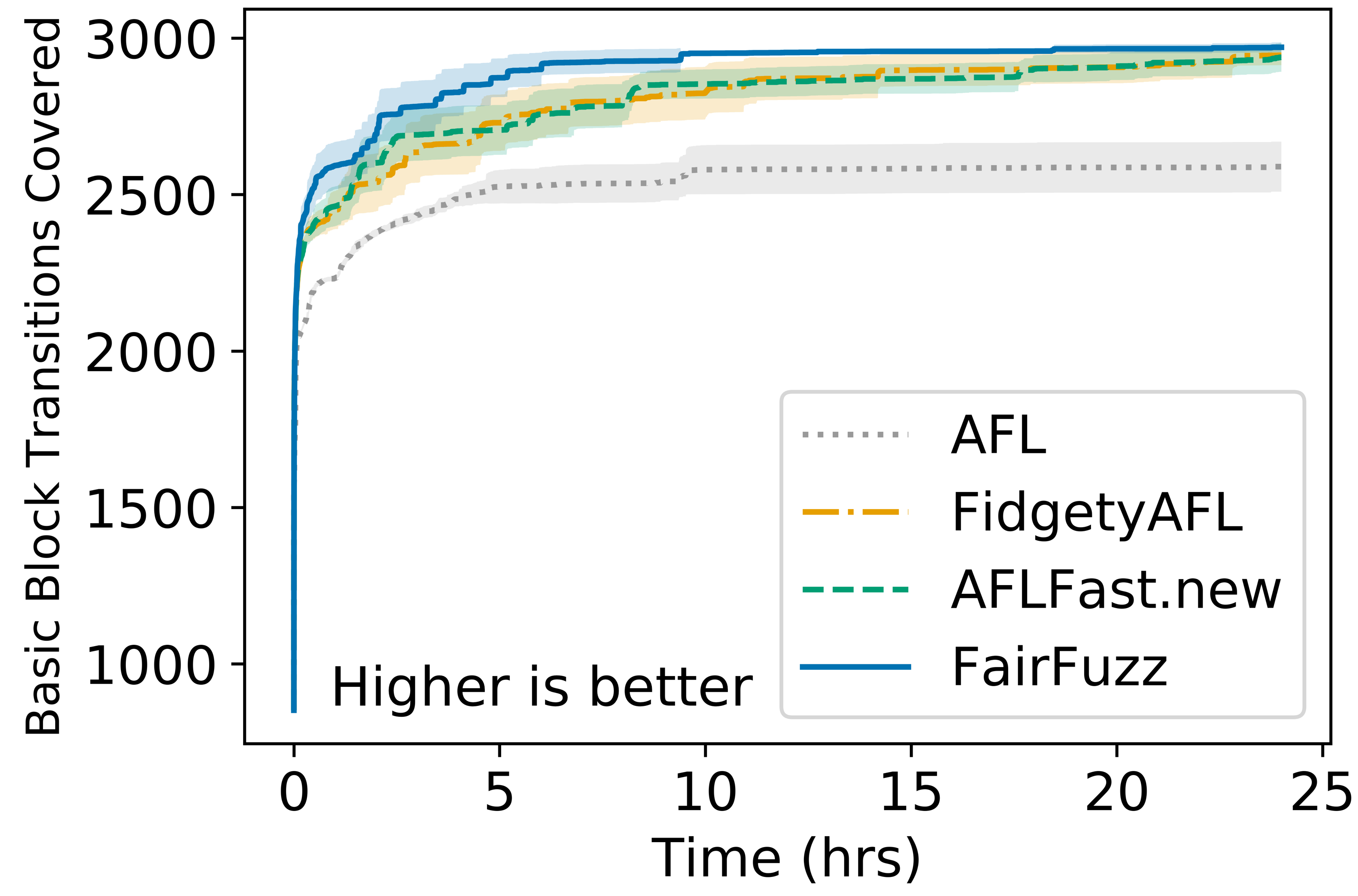}
\end{subfigure}
\caption{Number of basic block transitions (AFL branches) covered by different AFL techniques averaged over 20 runs (bands represent 95\% C.I.s). } \label{fig:coverage-vs}
\end{figure*}

In this evaluation we compare three popular versions of AFL against
\lfb{}, all based off of AFL version 2.40b.  ``AFL'' is the vanilla
AFL available from AFL's website.  ``FidgetyAFL''~\cite{FidgetyAFL} is
AFL run without the deterministic mutation stage, which AFL's creator
found replicated the performance of AFLFast~\cite{Bohme16} on some
benchmarks.  ``AFLFast.new''~\cite{AFLFast.new} is AFLFast run with
new settings, which has been claimed as significantly better than both
FidgetyAFL and AFLFast.  AFLFast.new omits the deterministic stage and
uses the cut-off exponential exploration strategy.
We ran \toolname{} with input trimming for the target branch and
omitting all deterministic stages except those necessary for the
creation of a branch mask. Our initial experiments to determine which
of our modifications (trimming, using the branch mask, doing all
deterministic mutations) were most effective were
inconclusive. 
This combination was a compromise which saw coverage increases on
multiple
benchmarks. 
We will refer to techniques (2), (3) and (4) as the ``modified''
techniques.

To evaluate their ability to achieve fast code coverage and discover
crashes, we evaluated the techniques on 9 different benchmarks. We
selected these from those favored for evaluation by the AFL creator
(\djpeg{} from libjpeg-turbo-1.5.1, and \readpng{} from
libpng-1.6.29), those used in AFLFast's evaluation (\texttt{tcpdump -nr} from tcpdump-4.9.0; and
\nm{}, \texttt{objdump -d}, \texttt{readelf -a}, and \cxxfilt{} from GNU
binutils-2.28) and a few benchmarks with more complex input
grammars in which AFL has previously found vulnerabilities (\texttt{mutool draw}
from mupdf-1.9, 
and \xmllint{} from libxml2-2.94). 
Since some of these input formats had AFL dictionaries and some did
not, we ran all this evaluation without dictionaries to level out the
playing field. In each case we seeded the fuzzing run with the inputs
in the corresponding AFL \texttt{testcases} directories (except \cxxfilt{}, which was seeded with the input ``\texttt{\_Z1fv\textcolor{violet}{\textbackslash n}}'').
We ran each technique for 24 hours (on a single core) on each benchmark, repeating each
24 hour run 20 times for each
benchmark. 

The main metric we report is basic block transitions
covered, which is close to the notion of branch coverage used in
real-world software testing.  Due to AFL's implementation, of the three common metrics used to
evaluate versions of AFL---number of paths, number of unique crashes,
and number of basic block transitions covered---this is the only one
that is robust to the order in which inputs are
discovered. 
As a simple illustration, 
consider a program with two branches, $b_1$ and $b_2$. Suppose input
$A$ hits $b_1$ once, input $B$ hits $b_2$ once, and input $C$ hits
both $b_1$ and $b_2$. Their respective paths are $p_A=\{(b_1,1)\}$,
$p_B=\{(b_2,1)\}$, and $p_C=\{(b_1,1), (b_2,1)\}$. Now, if AFL
discovers these inputs in the order $A,B,C$ it will save both $A$ and
$B$ and count 2 paths (or 2 unique crashes if both crash), and not
save $C$ since it does not exercise a new (\textit{branch ID},
\textit{hit count}) pair. On the other hand, if AFL discovers the
inputs in the order $C,A,B$, it will save $C$ and count 1 path (or 1
unique crash if $C$ crashes), and save neither $A$ nor $B$. Thus it
appears on the second run that AFL has found half the paths it did on the first run (or unique
crashes). On the other hand, regardless of the order in which inputs
$A,B,C$ are discovered, the number of basic block transitions covered
will be 2. We note that the creator of AFL also favors basic block
transitions covered over unique crashes as a performance
metric~\cite{no-unique-crashes}. We will nonetheless present data on
AFL-measured unique crashes in Section~\ref{sec:eval-crashes} for
comparison to prior work.

Although previous work does not always repeat
experiments~\cite{Stephens16,Li17,Bohme17} or, when the experiments
repeated, provide any measure of variability for metrics like unique
crashes~\cite{Bohme16}, we found compelling evidence (see the results
in Sections~\ref{sec:eval-coverage} and~\ref{sec:eval-crashes} and in
particular the graphs in Figure~\ref{fig:coverage-vs} and
Figure~\ref{fig:unique-crashes}) of the importance of reporting the
variability of AFL results in some form or another. We repeat our
experiments 20 times because AFL is an inherently non-deterministic
process (especially when running only havoc mutations), and so is its
performance.  This enabled us to report results that are statistically
significant.  We believe that all future research work should perform
experiments along similar lines.


\subsection{Coverage Compared to Prior Techniques}
\label{sec:eval-coverage}

In analyzing the program coverage achieved by each technique we sought
to answer two research questions:
\begin{enumerate}
\item[RQ1] Does an emphasis on rare branches result in faster program coverage?
\item[RQ2] Does an emphasis on rare branches lead to long-term gains in coverage?
\end{enumerate}

To answer these question we begin by analyzing coverage as measured in the number of basic block transitions (branches) discovered
through time.  Figure~\ref{fig:coverage-vs} plots, for each benchmark
and technique, the average number of branches covered over all 20 runs
at each time point (dark central line) and 95\% confidence intervals
in branches covered at each time point (shaded region around line)
over the 20 runs for each benchmark. For the confidence intervals we
assume a Student's t Distribution (taking $2.0860$ times the standard
error).

\begin{figure}
\includegraphics[width=\columnwidth]{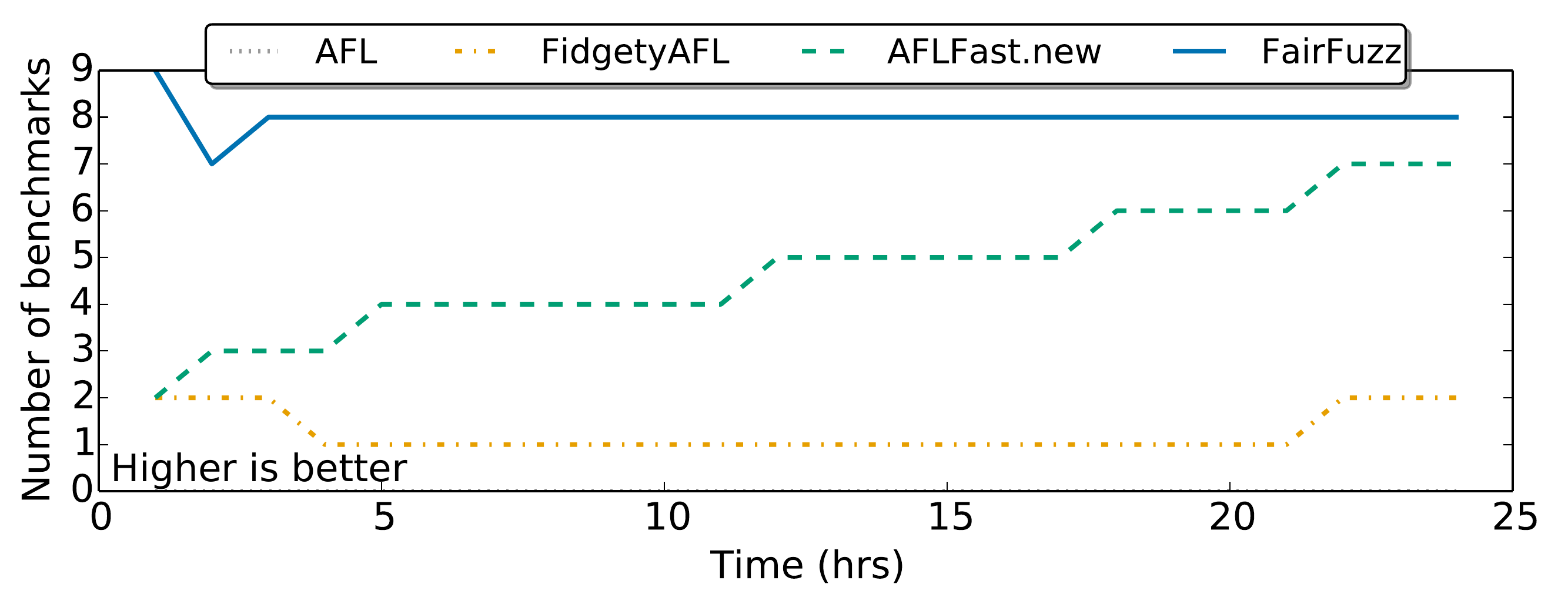}
\caption{Number of benchmarks on which each technique has the lead in coverage at each hour. A benchmark is counted for multiple techniques if two techniques are tied for the lead.}\label{fig:coverage-leaders}
\end{figure}

We can see in this Figure~\ref{fig:coverage-vs} that on most benchmarks, \toolname{}
achieves the upper bound in branch coverage, generally showing the
most rapid increase in coverage at the beginning of execution.
Overall, \lfb{} produces rapid increases in coverage compared
to other techniques on \objdump{}, \readelf{}, \readpng{}, \tcpdump{},
and \xmllint{}, and to a lesser degree on \nm{}, while tying
with the other techniques on \mutool{} and \djpeg{}, and AFLFast.new
having the edge on \cxxfilt{}.  Note that while \lfb{} keeps a sizeable lead on the \xmllint{} benchmark (Figure~\ref{fig:coverage-xmllint}), it does so wide variability. Closer analysis reveals that one run of \lfb{} on \xmllint{} revealed a bug in the 
rare branches queueing strategy, causing \lfb{} to select no inputs for mutation---this run covered no more than 6160 branches. However, \lfb{} had two runs on \xmllint{}
covering an exceptional 7969 and 10990 branches, respectively. But even without these three outliers, \lfb{}'s average coverage (6896) remains much higher than the other techniques' averages ( AFLFast.new
averages 6541 branches after 24 hours).

Figure ~\ref{fig:coverage-leaders},
shows, at every hour, for how many benchmarks each technique has
the \emph{lead} in coverage. By \emph{lead} we mean its average coverage is above the
 confidence intervals of the other techniques, and no other technique's
average lies within its confidence interval. We say two techniques are
tied if one's average lies within the confidence interval of the
other. If techniques tie for the lead, the benchmark is counted for both techniques in Figure~\ref{fig:coverage-leaders}, which is why the number of benchmarks at each hour may add up to more than 9. This figure shows that \lfb{} quickly achieves a lead in coverage on nearly all benchmarks and is not surpassed in coverage by the other techniques in our time limits. 

We believe Figures~\ref{fig:coverage-vs} and~\ref{fig:coverage-leaders} are compelling evidence that in the context of our benchmarks, the answer to our first research question is \emph{yes}, an emphasis on rare
branches leads to faster program coverage.

%

\subsubsection{Detailed analysis of coverage differences.}
\label{sec:eval-code-coverage}

Figure~\ref{fig:coverage-vs} shows there are three benchmarks
(\cxxfilt{}, \tcpdump{}, and \xmllint{}) on which one technique
achieves a lead in AFL's branch coverage after 24 hours (with AFLFast.new leading on \cxxfilt{} and \lfb{} on the other two). So, to answer our second research question, we do a more in depth analysis of the coverage achieved. In particular, we examine what the difference in AFL branch coverage corresponds to in terms of source code coverage.

Since AFL saves all inputs that achieve new program coverage (i.e. that 
are placed in the queue) to disk, we can replicate what program 
coverage was achieved in each run
by replaying these queue elements through the programs under test.
Since each benchmark was run 20 times, we take the union (over each
technique) of inputs in the queue for all 20 runs.  We ran the union
of the inputs for each technique through their corresponding programs
and then ran \texttt{lcov} on the results to reveal coverage differences.

\paragraph{\xmllint{}}

The bulk of the coverage gains on \xmllint{} were in the main \texttt{parser.c}
file. The key trend in increased coverage appears to be \lfb{}'s
increased ability to discover keywords.  For example, both AFL and
\lfb{} have higher coverage than FidgetyAFL and AFLFast.new as they
discovered the patterns \texttt{<!DOCTYPE} and \texttt{<!ATTLIST} in
at least one run.  However, \lfb{} also discovered the
\texttt{\#REQUIRED}, \texttt{\#IMPLIED}, and \texttt{\#FIXED}
keywords, producing inputs including:
\begin{center}
\texttt{<!DOCTYPE6[ <!ATTLIST\'i T ID \#FIXED\%}\\
\texttt{<!DOCTYPE\brokenvert[ <!ATTLIST\'iD T ID \#IMPLIEDOCTY}\\
\texttt{<!DOCTYPE\textbackslash[ <!ATTLIST\'iD T ID \#REQUIRED\textcolor{violet}{\^{}@\^{}P}}\\
\end{center}

We found several other instances of keywords discovered by Rare
Branches but not the other techniques for \xmllint{}. We believe our
rare branch targeting technique is directly responsible for this. To see this,
let us focus on the \texttt{<!ATTLIST>} block covered by the inputs
above, whose code is outlined in in Figure~\ref{fig:xmllint-ex}. While
both AFL and \lfb{} had a run discovering the sequence
\texttt{<!ATTLIST}, running \xmllint{} on all the saved inputs produced by
AFL only resulted in 18 hits of Line 2 of Figure~\ref{fig:xmllint-ex}. Running \xmllint{}
on all saved inputs produced by \lfb{}, on the other hand, resulted in
\emph{2124} hits of this line. With the queued inputs resulting in two orders of magnitude more hits of this line, it is obvious that \lfb{} was
better able to discover inputs with more structure than AFL.

It is valid to ask whether this increase is only attributable to luck,
since after all, it was only in one run that AFL and \lfb{}
discovered the full \texttt{<!ATTLIST} sequence. While we believe the
believe the difference in the number of inputs hitting Line 2 of
Figure~\ref{fig:xmllint-ex} for these two techniques is compelling
evidence this was not the case, looking at the number of runs which
produced subsequences of \texttt{<!ATTLIST} also seems to suggest it
is not simply luck. As we can see in Table~\ref{table:keyword-progression}, the decrease in the number of runs discovering
\texttt{<!AT} from the number of runs discovering \texttt{<!A} 
shows the branch mask in action, with 11 of \lfb{}' runs
discovering \texttt{<!AT}, compared to 1, 2, and 3 for AFLFast.new,
FidgetyAFL, and AFLFast.new, respectively.

\begin{table}
\begin{tabular} {| l | r r r r|}
\hline
subsequence & AFL & FidgetyAFL & AFLFast.new & \lfb{} \\ 
\hline
\texttt{<!A} & 7 & 15 & 18 & 17\\
\texttt{<!AT} &1 & 2 & 3 & 11\\
\texttt{<!ATT} &1 & 0 & 0& 1\\
\hline
\end{tabular}
\caption{Number of runs, for each technique, producing a seed input with the given subsequence in 24 hours.}\label{table:keyword-progression}
\end{table}

We suspect \lfb{} performs particularly well on this benchmark
due to the structure of the code in \texttt{parser.c}. The
\texttt{CMP}$n$ macros in this code (see Figure~\ref{fig:xmllint-ex})
expand into character-by-character string comparisons. AFL is
able to explore these more easily than full string comparisons since this splits the comparison into different basic blocks, progress through which is reported as new coverage. 
We discuss a recently proposed extension to the LLVM version of the AFL instrumenter~\cite{laf-intel} that would automatically expand comparisons in programs not structured in
this manner in Section~\ref{sec:discussion}.

Finally, as is obvious from the example inputs above, although \lfb{} discovered more keywords, the inputs it produced were not necessarily more well-formed. Nonetheless,
these inputs allowed the \lfb{} to explore more of the
program's faults. This is reflected in the coverage of a large case
statement differentiating 57 error messages in \texttt{parser.c}. Both
FidgetyAFL and AFLFast.new cover only 22 of these cases, AFL covers 33,
and \lfb{} covers 39. Interestingly, \lfb{} misses one
error that the other algorithms cover---an error message about
excessive name length---indicating it did not produce an input with a
long enough name to trigger this error. We will come back to this when examining the coverage differences in \cxxfilt{}.

\paragraph{\tcpdump{}} 
We observed that coverage for \tcpdump{} differs a bit for all four techniques over a variety of different files. We see the biggest gains in three files printing certain packet types, all of which suggest \lfb{} is better able to automatically detect the structure of inputs.

In  \texttt{print-forces.c}, a ForCES Protocol printer (RFC 5810),  all of AFL, FidgetyAFL, and
AFLFast.new are able to create files that pass the first validity checks
on message type, but only \lfb{} is able to create
files that have legal ForCES packet length. In \texttt{print-llc.c}, an IEEE 802.2 Logical Link Control (LLC)
parser,  \lfb{} was be able to create packets
with the organizationally unique identifier (OUI) corresponding to RFC 2684, and subsequently explore the
many subtypes of this OUI. Finally, in \texttt{print-snmp.c},  a Simple Network Management Protocol printer, we find that \lfb{} produces inputs corresponding to Trap
PDUs 
with valid timestamps while AFL does not. \lfb{} also
produces 
scoped SNMP PDUs with valid \texttt{contextName} while AFL does
not. Finally, in a function to decode a SNMPv3 user-based security
message header \lfb{} produces a few inputs which pass through
this function successfully, while AFL does not produce any with a
valid \texttt{msgAuthoritativeEngineBoots}. 

We note these gains in coverage seem less impressive than those of \lfb{} on \xmllint{}.
We hypothesize this is because \lfb{}'s performance on \tcpdump{} may reflect consistently higher coverage of the program, but most of this increase was matched in at least one run of the other techniques. The coverage produced by running all inputs produced by each obscures on which proportion of runs a certain packet structure was discovered.  This hypothesis is confirmed by comparing the number of branches covered \emph{by at least one of} the 20 runs (the union over the runs) and the number of branches covered \emph{at least once in all} the 20 runs (the intersection over the runs) for the different techniques.  For \tcpdump{}, we see \lfb{} has a much more
consistent increase in coverage (the intersection of coverage for
\lfb{} contains 11,293 branches, 500 more branches than AFLFast.new's
10,724), but no huge standout gain the union (the union of coverage is 16,129, only 200 more branches than
AFLFast.new's 15,929). \lfb{}'s performance on \xmllint{} shows the opposite behavior.  The intersection of coverage for \xmllint{}
is virtually the same for the three modified techniques (5,876 for
FidgetyAFL, 5,778 for AFLFast.new and 5,884 for \toolname{}), but
\toolname{}'s union of coverage (11,681) contains over 4,000 more
branches than AFLFast.new's union of coverage (7,222). 

\paragraph{\cxxfilt{}} 
The differences in terms of source code coverage between techniques were much more minimal for \cxxfilt{} than for \tcpdump{} or \xmllint{}.
We list all the differences in coverage between AFLFast.new and \lfb{} below. \lfb{}'s gains are structural in the input type; it
covers 3 lines in \texttt{cp-demangle.c} that AFLFast.new does
not, 
related to demangling binary components when the operator has a
certain opcode.  AFLFast.new's gains appear to be mostly related to
\lfb{}'s inability to produce very long inputs. In
\texttt{cp-demangle.c}, AFLFast.new covers a branch where
\texttt{xmalloc\_failed(INT\_MAX)} is called if a length bound
comparison fails. \lfb{} fails to produce an input long enough
to violate the length bound. \lfb{} also fails to cover a
branch in \texttt{cxxfilt.c}, taken when the length of input read into
\cxxfilt{} surpasses the length of the input buffer allocated to store
it, which all other techniques cover.

\lfb{}'s inability to produce very long inputs may be due to the
second round of trimming \lfb{} does. It could also be due to the fact that a focus on
branches, ignoring hit counts an input achieves of a particular
branch, may not encourage \lfb{} to explore repeated iterations of
loops. 
In spite of the fact that this shows up as a small difference in
measurable coverage, \lfb{} shows very different behavior when
it comes to finding crashes, and we suspect the input length issue is
a reason why. We will discuss this in more detail in
Section~\ref{sec:eval-crashes}.

The pattern we see from this analysis is that \lfb{}  is
better able to automatically discover input constraints and
keywords---special sequences, packet lengths, organization codes---and
target exploration to inputs which satisfy these constraints than the
other techniques.  We suspect the gains in coverage speed on benchmarks such as \objdump{}, \readpng{}, and \readelf{} are due to similar factors. We conjecture the
targeting of rare branches shines the most in the \tcpdump{} and
\xmllint{} benchmarks since these programs are structured with many
nested constraints, which the other techniques are unable to properly
explore over the time budget (and perhaps even longer) without extreme luck.

\subsection{Are branches successfully targeted?}
\label{sec:eval-targeting-branches}

Our third research question focused on the effectiveness of the branch mask strategy, i.e.:
\begin{enumerate}
\item[RQ3] Does the branch mask increase the proportion of produced inputs hitting the target branch?
\end{enumerate}

To evaluate this we conducted an experiment where, for each seed input chosen for mutation we first ran a \emph{shadow run} of mutations with the branch mask disabled, then re-ran the mutations with the branch mask enabled. We also disabled side effects (queueing of inputs, incrementing of branch hit counts) in the shadow run.  This shadow run allows us to compute the difference between the percentage of generated inputs hitting the branch with and without the branch mask \emph{for each seed input}. 

We ran \lfb{} with these shadow runs for one queueing cycle on a subset of our benchmarks. 
For each benchmark we ran a cycle with target branch trimming and one without. Table~\ref{table:shadow-mode-results} shows the target branch hit percentage for the deterministic\footnote{We separated the byteflipping stage in which the branch mask is computed--- the \emph{calibration stage}---from the subsequent deterministic stages which use the branch mask. 
} and havoc stages, averaging the per-input hit percentages over all inputs generated in the first cycle. 

\begin{table}
\caption{Average \% of mutated inputs hitting target branch for one queueing cycle.} 
\label{table:shadow-mode-results}
\begin{subtable}{\columnwidth}
\caption{Cycle without trimming.}\label{table:shadow-mode-notrim}
\begin{tabular}{|l | c | c| c| c| c| c|}
\hline
 & det. mask & det plain & havoc  mask & havoc plain\\ \hline
\xmllint{}   & 92.8\%&46.5\%& 31.8\% & 6.6\%\\
\tcpdump {} & 99.0\%  &  74.0\% & 34.2\% &9.3\%\\
\cxxfilt {} & 97.6\%  & 64.1\% & 41.4\%&14.4\%\\
\readelf{} &99.7\% &82.7\%& 57.7\%& 14.9\%\\
\readpng{}  & 99.1\% & 34.6\%& 24.3\%& 2.4\% \\
\objdump{}  & 99.2\%  & 70.2\%& 42.4\% & 9.0\%\\
\hline
\end{tabular}
\end{subtable}%

\begin{subtable}{\columnwidth}
\caption{Cycle with trimming.}\label{table:shadow-mode-trim}
\begin{tabular}{|l | c | c| c| c| c| c|}
\hline
 & det. mask & det. plain & havoc  mask & havoc plain\\ \hline
\xmllint{} &90.3\%& 22.9\%& 32.8\%& 2.9\%\\
\tcpdump {} & 98.7\% & 72.8\% & 36.1\% & 9.0\% \\
\cxxfilt {} & 96.6\% &  14.8\% & 34.4\% & 1.1\%\\
\readelf{}& 99.7\% & 78.2\%& 55.5\% & 11.4\%\\
\readpng{} &97.8\%  &39.0\%  & 24.0\%& 2.4\% \\
\objdump{} &99.2\% & 66.7\% & 46.2\% &7.6\% \\
\hline
\end{tabular}
\end{subtable}
\end{table}

Comparing the numbers in Table~\ref{table:shadow-mode-notrim} and Table~\ref{table:shadow-mode-trim}, it appears that trimming the input reduces the number of inputs hitting the target branch when the branch mask is disabled but has minimal effect when the branch mask is enabled. Overall, Table~\ref{table:shadow-mode-results} shows that the branch mask does largely increase the percentage of mutated inputs hitting the target branch. The hit percentages for the deterministic stage are strikingly high: this is not unexpected as in the deterministic stage the branch mask simply operates to prevent mutations at locations likely to violate the target branch. 
What is most impressive is the gain in the percentage of inputs hitting the target branch in the havoc stage. 
In spite of the use of the branch mask in the havoc stage being approximate, we consistently see the use of the branch mask causing a 3x-10x increase in the percentage of inputs hitting the target branch. 
From this analysis, we conclude that the answer to our third research question is positive -- the branch mask increases the proportion of inputs hitting the target branch.

Note that this branch mask targeting technique is independent of the fact that the branches being targeted are ``rare''. This means that this strategy could be used in a more general context. For example, we could target only the branches within a function that needs to be tested, or, if some area of the code was recently modified or bug-prone, we could target the branches in that area with the branch mask. Recent work on targeted AFL~\cite{Bohme17} shows promise in such an application of AFL, and we believe the branch mask technique could be used cooperatively with the power schedules presented in this work.

\subsection{Crashing Compared to Prior Techniques}
\label{sec:eval-crashes}

\begin{figure}
\begin{subfigure}[t]{0.5\columnwidth}
\includegraphics[width=\columnwidth]{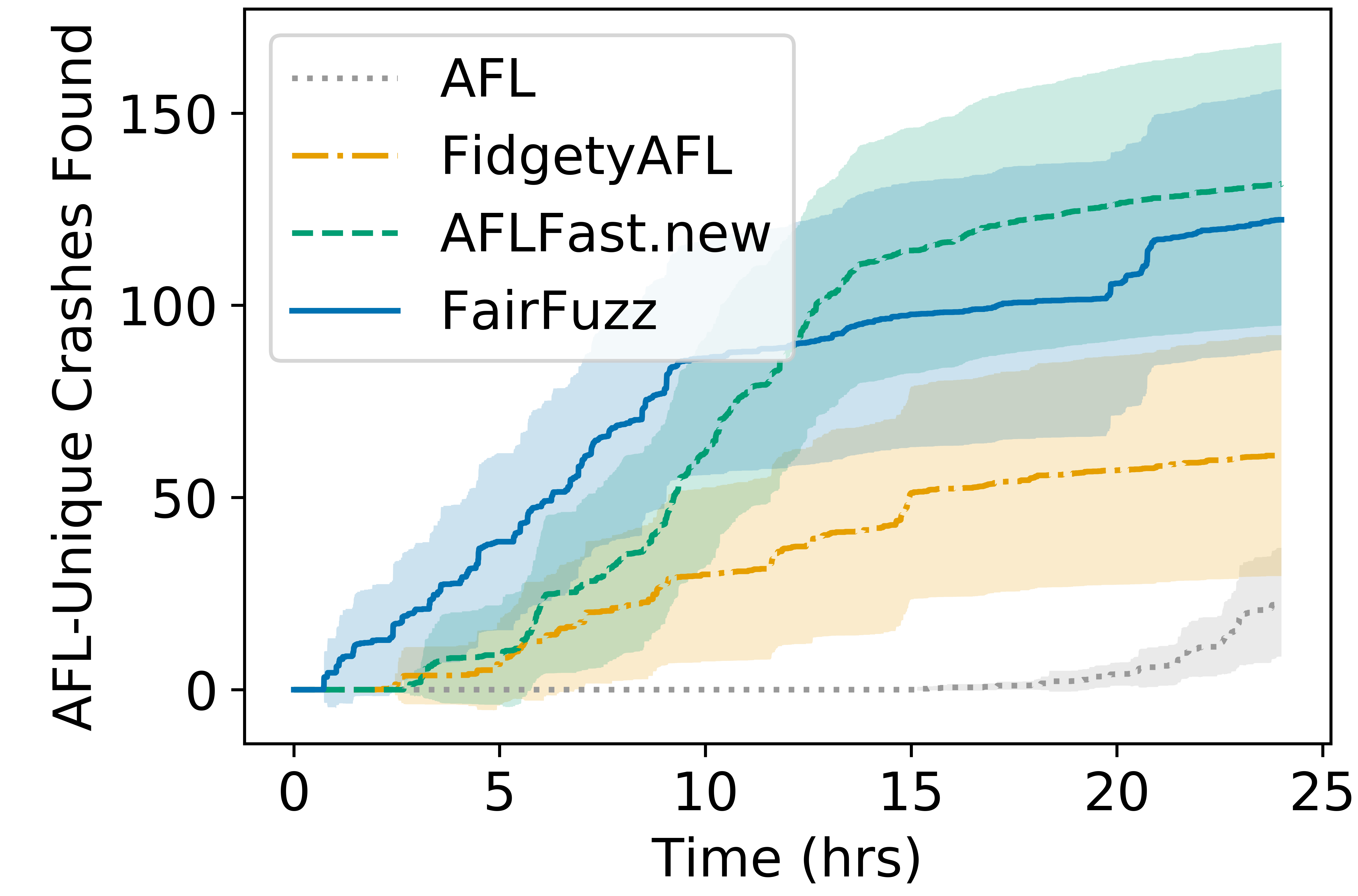}
\subcaption{readelf}\label{fig:unique-crashes-readelf}
\end{subfigure}%
\begin{subfigure}[t]{0.5\columnwidth}
\includegraphics[width=\columnwidth]{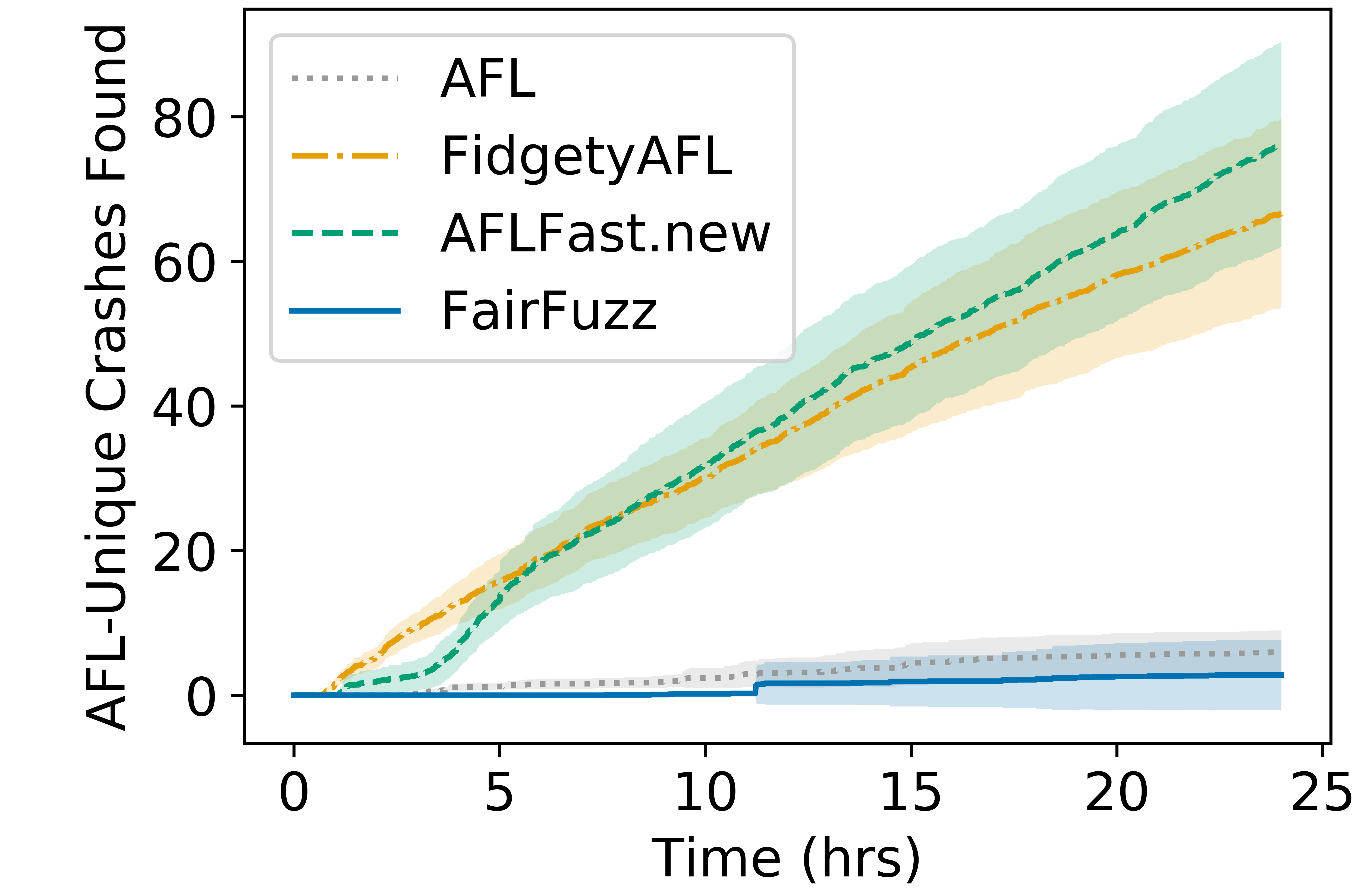}
\subcaption{cxxfilt}\label{fig:unique-crashes-cxxfilt}
\end{subfigure}
\caption{Average unique crashes found over 20 runs for each technique (dark line) with 95\% confidence intervals.}\label{fig:unique-crashes}
\end{figure}

Our final research question pertains to crash finding:
\begin{enumerate}
\item[RQ3.] Does an emphasis on rare branches lead to faster crash exposure?
\end{enumerate}
Crash finding is hard to evaluate in practice. Of the 9 benchmarks we tested on, all techniques found crashes only on \cxxfilt{} and \readelf{}. 
While we discussed issues with the unique crashes metric at the beginning of this section, for comparison with prior work, we plot the unique crashes found by each technique in Figure~\ref{fig:unique-crashes}. The most interesting part of this figure is the width of the confidence intervals, which reflects just how variable this metric is. 

\begin{figure}
\raggedright
\begin{subfigure}[t]{0.95\columnwidth}
\subcaption{readelf}\label{fig:crashes-readelf}
\includegraphics[width=\columnwidth]{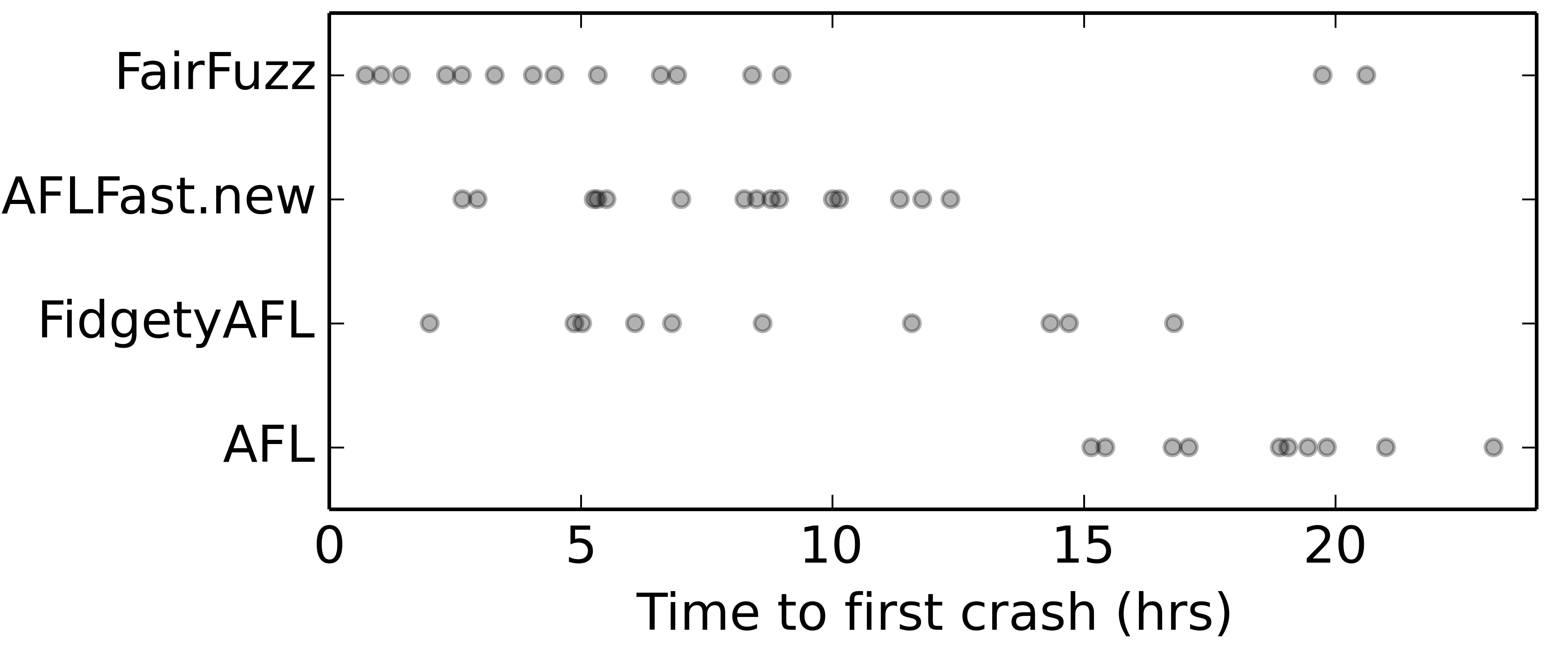}
\end{subfigure}
\begin{subfigure}[t]{0.95\columnwidth}
\subcaption{cxxfilt}\label{fig:crashes-cxxfilt}
\includegraphics[width=\columnwidth]{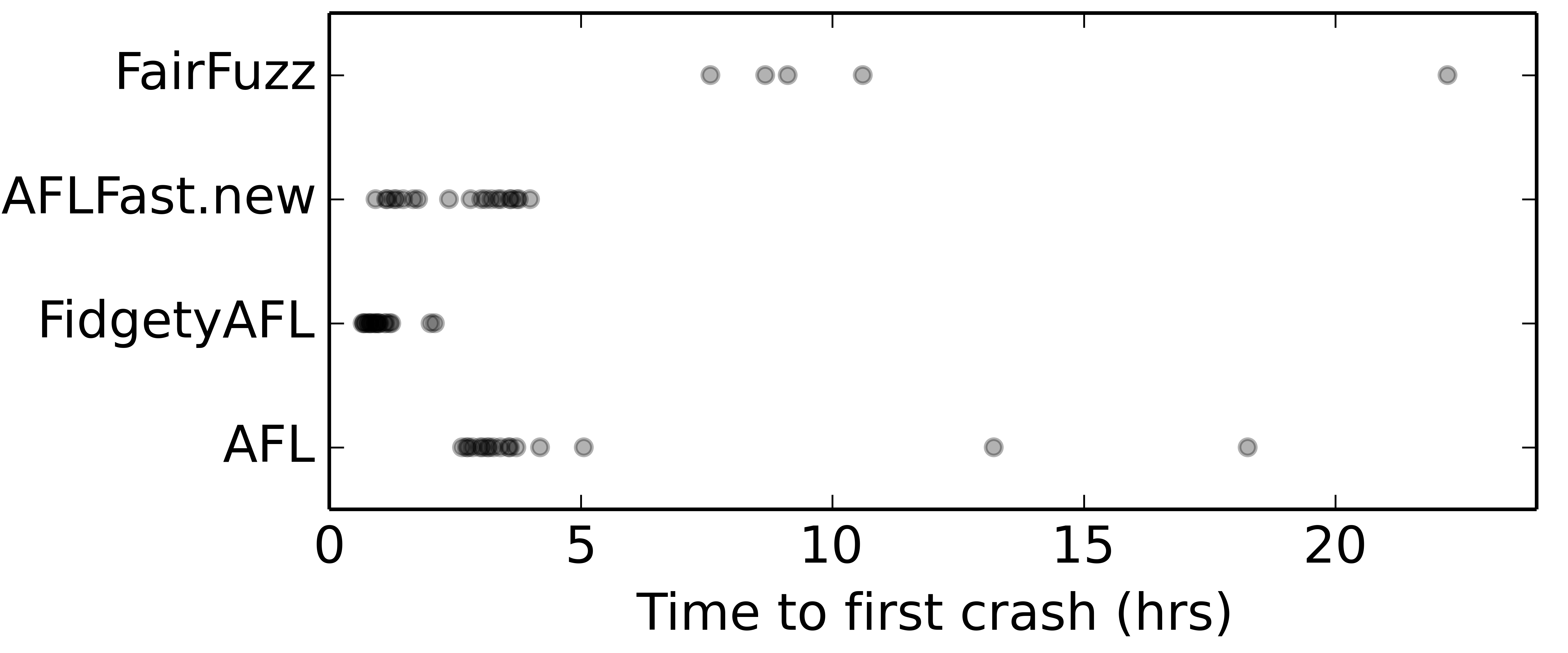}
\end{subfigure}
\caption{Time to find first crash for each run of the different techniques. Each point represents the time to first crash for a single run.}\label{fig:crash-times}
\end{figure}

In spite of the variability, the figure suggests \lfb{} performs poorly on \cxxfilt{} in terms of crash exposure. To evaluate this, we can first compare the percent of runs finding crashes for each benchmark. On \readelf{}, both AFL and FidgetyAFL find crashes in 50\% of runs while AFLFast.new and \lfb{} find crashes in 75\% of runs. On \cxxfilt{}, however,  FidgetyAFL and AFLFast.new find crashes in 100\% of the runs, AFL in 85\% of them, and \lfb{} in only 25\%. Looking at the time to first crash for the different runs, we see that for \readelf{}, in general, \lfb{} finds crashes a bit faster than AFLFast.new (Figure~\ref{fig:crashes-readelf}). However, for the few runs on which the \lfb{} found crashes for \cxxfilt{}, it takes around 10 hours for \lfb{} to find the crashes as compared to less than 2 hours for FidgetyAFL. 



Given that some of the lines missed by \lfb{} in \cxxfilt{} were about buffer length (recall Section~\ref{sec:eval-code-coverage}), we suspected that input length might be a factor in this performance discrepancy. 
We examined the file size of the crashing inputs saved for each technique. Many of these were very long, more than 20KB, with one crashing input generated by AFLFast.new being 130KB long! We saw crashing inputs of this length in only one of the five \lfb{} runs which found crashes in \cxxfilt{}.

Thus, we suspect a factor in \lfb{}'s poor performance on crash finding on this \cxxfilt{} is that the method does not encourage the creation of huge inputs. It is also possible that the structure of \cxxfilt{} may is better suited to a path-based exploration strategy (like that of AFL/FidgetyAFL and AFLFast.new) than a branch-based one. For example, demangling function argument types is done in a loop, so \lfb{} is less likely to prioritize many new function arguments than path-based prioritization does. Overall, the answer to our final research question is that \lfb{} may improve crash exposure in some cases (\readelf{}), but does not when the crashes are most easily exposed by large inputs.

\section{Discussion}
\label{sec:discussion}

The foremost limitation of \toolname{} is the fact that branches that
are never hit by any AFL input cannot be targeted by this method. So,
it confers little benefits to discovering a single long magic number
(like that in Figure~\ref{fig:magic-number}) when progress towards
matching the magic number does not result in new coverage. However,
recall  \toolname{} was effective at finding keyword sequences in the \xmllint{} benchmark. This is because the long
string comparisons in \texttt{parser.c} were structured as
byte-by-byte comparisons, so AFL's instrumentation reported new
coverage when progress was made on these comparisons. AFL's
instrumentation would not report new coverage for progress on a single
multi-byte comparison.


The creators of \texttt{laf-intel}~\cite{laf-intel} propose several
LLVM ``deoptimization'' passes to improve AFL's performance, including
a pass that automatically turns multi-byte comparisons into
byte-by-byte comparisons. Figure~\ref{fig:laf-intel-ex} shows an
example of this comparison unrolling. The integration of these LLVM
passes into AFL's instrumentation is straightforward, requiring only a
patch to AFL's LLVM-based instrumenter~\cite{laf-intel-code}. Due to \toolname{}'s performance on the \xmllint{} benchmark, we
believe \toolname{} could show similar coverage gains on other
programs if they were compiled with this \texttt{laf-intel} pass. We
did not evaluate this as the evaluation of the \texttt{laf-intel} pass
was done in AFL's ``parallel'' fuzzing mode, with some instances of
AFL running on the traditionally instrumented program and others
running on the enhanced instrumentation programs. 
We did not do any experiments with this parallel fuzzing as our
implementation did not implement a distributed version of our rare
branch computation algorithm.  

\begin{figure}
\begin{minipage}[t]{0.45\columnwidth}
\vspace{0.5cm}
\begin{lstlisting}[language=customC]
if ( str == "BAD!") {
   // do bad things
}
\end{lstlisting}
\end{minipage}
\begin{minipage}[t]{0.45\columnwidth}
\begin{lstlisting}[language=customC]
if (str[0]  == 'B') {
   if (str[1] == 'A') {
      if (str[2] == 'D') {
         if (str[3] == '!') {
             // do bad things
}}}}
\end{lstlisting}
\end{minipage}
\caption{Multi-byte comparison (left) unrolled to byte-by-byte comparison (right).}\label{fig:laf-intel-ex}
\end{figure}

\section{Other Related Work}

Unlike \lfb{} and other greybox fuzzers~\cite{libFuzzer} which use coverage information as a heuristic for which inputs may yield new coverage under mutation, symbolic execution tools~\cite{Godefroid05,Sen06,Cadar08, Sen05} methodically explore the program under test by capturing path constraints and directly producing inputs which fit yet-unexplored path constraints. This cost of this precision is that it can lead to the path explosion problem, which causes scalability issues.

Traditional blackbox fuzzers such as zzuf~\cite{Hocevar07} mutate user-provided seed inputs according to a mutation ratio, which may need to be adjusted to the program under test. BFF~\cite{Householder12} and \textsc{SymFuzz}~\cite{Cha15} adapt this parameter automatically, by measuring crash density (number of mutated inputs finding crashes) and doing input bit dependence, respectively. These optimizations are not relevant to AFL-type generational fuzzers which do not use this mutation ratio parameter.

There exist several fuzzers highly optimized for certain input file structures, including network protocols~\cite{Amini07,Bratus08}, and source code~\cite{Yang11,Holler12,Ruderman15}. \lfb{} is of much lower specificity so will not be as effective as these tools on these specific input formats. However, its method is fully automatic, requiring neither user inputs~\cite{Amini07} or extensive tuning~\cite{Yang11,Ruderman15}.

While \lfb{} uses its branch mask to try and fix important parts of program
inputs, recent work has more explicitly tried to automatically learn input formats. Learn\&Fuzz~\cite{Godefroid17} uses sequence-based learning methods to learn the structure of pdf objects, leveraging this to produce new valid pdf. \textsc{Autogram}~\cite{Hoschele16} proposes a taint analysis- based approach to learning input grammars, while \textsc{Glade}~\cite{Bastani17} uses an iterative approach and repeated calls to an oracle to learn a context-free grammar for a set of inputs. We also do not assume a corpus of valid inputs of any size from which validity could be automatically learned in our work. 


Another approach to smarter fuzzing is to find locations in seed inputs related to likely crash locations in the program and focus mutation there. BuzzFuzz~\cite{Ganesh09} and  TaintScope~\cite{Wang10} pair taint analysis with the detection of potential attack points (i.e. array allocations) to find crashes more effectively. TaintScope also includes automated checksum check detection. Dowser~\cite{Haller13} has a similar goal but instead guides symbolic execution down branches likely to find buffer overflows. These directed methods are not directly comparable to \lfb{} since they do not have the same goal of broadening and deepening program exploration. 

VUzzer's~\cite{Rawat17} approach is reminiscent of these, using both static and dynamic analysis to get immediate values used in comparisons (likely to be magic numbers), positions in the input likely to be magic numbers are offset, and information about which basic blocks are hard to get to. It models the program as a Markov Chain to decide with parts of the program are rare, as opposed to our empirical approach.

Randoop~\cite{Pacheco07} automatically generates test cases for object oriented programs through feedback-directed random test generation, checking to make sure program fragments are valid; Evosuite~\cite{Fraser11} uses seed inputs and genetic
algorithms to achieve high code coverage. Both these techniques focus on generating sequences of method calls to test programs, which is quite different from the input-generation form of fuzzing in \lfb{}.

Search-based software testing (SBST)~\cite{miller1976automatic,korel1990automated,McMinn-SBST,grechanik2009maintaining,grechanik2012automatically,harman2001search,harman2007current,harman2004metrics,yoo2007pareto}
uses optimization techniques such as hill climbing and
genetic algorithms to generate inputs that optimize some observable
fitness function.  These techniques work well when the fitness curve
is smooth with respect to changes in the input, which is not the case in coverage-based greybox fuzzing.

\bibliographystyle{ACM-Reference-Format}
\bibliography{lfb}

\end{document}